\PassOptionsToPackage{square,comma,numbers,sort&compress,super}{natbib}
\documentclass[10pt,twocolumn,twoside]{pnas-new}
\usepackage{balance} 
\usepackage{fancyvrb}

\usepackage{adjustbox}

\usepackage{stackengine,scalerel}

\templatetype{pnasresearcharticle} 

\usepackage{setspace}

\usepackage{booktabs}
\usepackage{subfig}
\usepackage[11pt]{moresize}
\renewcommand{\marginpar}[1]{}

\usepackage{mathtools}
\usepackage{times}
\usepackage{urcsbiblio}
\usepackage{latexsym}
\usepackage{multirow}
\usepackage{arydshln}

\usepackage{graphicx}
\usepackage{ulem}	
\usepackage{examples}
\usepackage{pdfpages}
\usepackage{xspace}
\usepackage{amssymb,amsmath,epsfig}
\usepackage{mathpartir}
\usepackage{amsthm}
\usepackage{mathrsfs}
\usepackage{algorithm}
\usepackage{pifont}
\usepackage[noend]{algpseudocode}
\usetikzlibrary{matrix, positioning, arrows.meta, calc, shapes, decorations,
  backgrounds, arrows, decorations.pathreplacing}
\usepackage{pgfplots, siunitx}
\usepackage{hhline}

\usepackage{soul}

\newcommand{\nts}{\ensuremath{\text{\it nt}}\xspace}

\newcommand{\notes}[1]{}

\newcommand{\ith}[1]{\ensuremath{i^{{th}}}}

\newcount\permx
\newcount\permy
\def\permdot#1#2{
\permx=#1 \advance\permx by-1
\permy=#2 \advance\permy by-1
\psframe[fillcolor=black, fillstyle=solid]
(\permx,\permy)(#1, #2)
}

\newcommand{\argmin}{\operatornamewithlimits{\mathbf{argmin}}}

\newcommand{\vecx}{\ensuremath{\mathbf{x}}\xspace}
\newcommand{\vecy}{\ensuremath{\mathbf{y}}\xspace}
\newcommand{\vecw}{\ensuremath{\mathbf{w}}\xspace}

\newcommand{\vecyhat}{\ensuremath{\hat{\vecy}}\xspace}

\newcommand{\smallnt}[1]{\ensuremath{_{\mbox{\tiny PP}}}\xspace}

\iffalse

\else

\fi

\newcommand{\smallurl}[1]{{\scriptsize \url{#1}}}

\newcommand{\pop}{\ensuremath{\mathsf{pop}}\xspace}

\newcommand{\nskip}{\ensuremath{\mathsf{skip}}\xspace}

\newcommand{\daromatica}{{\em D.~aromatica}\xspace}

\newcommand{\linearfold}{{LinearFold}\xspace}
\newcommand{\linearpartition}{{LinearPartition}\xspace}

\newcommand{\lcf}{{LinearCoFold}\xspace}
\newcommand{\lcp}{{LinearCoPartition}\xspace}
\newcommand{\linearcofold}{{LinearCoFold}\xspace}
\newcommand{\linearcopartition}{{LinearCoPartition}\xspace}

\newcommand{\viennarnacofold}{{Vienna RNAcofold}\xspace}
\newcommand{\rnacofold}{{RNAcofold}\xspace}
\newcommand{\pairfold}{{PairFold}\xspace}

\newcommand{\subscript}[2]{_{{#1}, {#2}}}
\newcommand{\Qf}[2]{\ensuremath{Q\subscript{{#1}}{{#2}}}\xspace}
\newcommand{\Qhat}{\ensuremath{\widehat{Q}}\xspace}
\newcommand{\Qhatf}[2]{\ensuremath{\Qhat\subscript{{#1}}{{#2}}}\xspace}
\newcommand{\pluseq}{\mathrel{+}=}

\newcommand{\Cf}[2]{\ensuremath{C\subscript{{#1}}{{#2}}}\xspace}

\newcommand{\panel}[1]{\large \sf {#1}}
\newcommand{\paneltext}[1]{\normalsize \sf {#1}}

\newcommand{\vecxa}{\ensuremath{\mathbf{x^a}}\xspace}
\newcommand{\vecxb}{\ensuremath{\mathbf{x^b}}\xspace}

\newcommand{\candidates}{\ensuremath{\mathit{candidates}}\xspace}

\newcommand{\statee}{\ensuremath{\mathbf{E}(i, j)}\xspace}
\newcommand{\statep}{\ensuremath{\mathbf{P}(i, j)}\xspace}
\newcommand{\statem}{\ensuremath{\mathbf{M}^{1}(i, j)}\xspace}
\newcommand{\statemm}{\ensuremath{\mathbf{M}^{2}(i, j)}\xspace}
\newcommand{\statemn}{\ensuremath{\mathbf{M}^{1}_{\text{nicked}(i, j)}}\xspace}
\newcommand{\statemmn}{\ensuremath{\mathbf{M}^{2}_{\text{nicked}(i, j)}}\xspace} 
\usepackage[superscript,biblabel]{cite}

\usepackage{bbm}

\title{\lcf and \lcp: Linear-Time Algorithms for Secondary Structure Prediction of  Interacting RNA molecules}

\author[b,a,$\dag$]{He Zhang}
\author[a,$\dag$]{Sizhen Li}
\author[a]{Liang Zhang}
\author[c,d,e]{David H.~Mathews}
\author[a,$\clubsuit$]{Liang Huang}

\affil[a]{School of Electrical Engineering ~\& Computer Science,
  Oregon State University, Corvallis, OR 97330, USA}
\affil[b]{Baidu Research USA, Sunnyvale, CA 94089, USA}
\affil[c]{Dept. of Biochemistry ~\& Biophysics}
\affil[d]{Center for RNA Biology}
\affil[e]{Dept. of Biostatistics ~\& Computational Biology, University of Rochester Medical Center, Rochester, NY 14642, USA}

\leadauthor{Zhang} 

\authordeclaration{The authors declare no conflict of interest.\\[-0.5cm]}

\authorcontributions{Author contributions: 
L.H.~conceived the idea and directed the project. 
L.H. and H.Z.~designed algorithms.
H.Z.~implemented the code.
D.H.M.~guided the evaluation 
that S.L., H.Z. and L.Z.~carried out. 
H.Z. and S.L.~wrote the manuscript; L.H. and D.H.M.~revised it.
}
\correspondingauthor{
  \textsuperscript{$\dag$}
  Equal contribution; 
  \textsuperscript{$\clubsuit$}
  corresponding author: {liang.huang.sh@gmail.com}. 
}

\begin{abstract}
Many ncRNAs function through RNA-RNA interactions.
Fast and reliable RNA structure prediction with consideration of RNA-RNA interaction is useful. 
Some existing tools are less accurate due to omitting the competing of intermolecular and intramolecular base pairs,
or focus more on predicting the binding region rather than predicting the complete secondary structure of two interacting strands. 
\viennarnacofold, which reduces the problem into the classical single sequence folding by concatenating two strands, 
scales in cubic time against the combined sequence length,
and is slow for long sequences.
To address these issues,
we present \lcf, which predicts the complete minimum free energy structure of two strands in linear runtime,
and \lcp,  which calculates the cofolding partition function and base pairing probabilities in linear runtime.
\lcf and \lcp follows the concatenation strategy of \rnacofold, but are orders of magnitude faster than \rnacofold.
For example, on a sequence pair with combined length of 26,190~\nts,
\lcf is $86.8 \times$ faster than \rnacofold MFE mode (0.6 minutes vs.~52.1 minutes),
and \lcp is $642.3 \times$ faster than \rnacofold partition function mode (1.8 minutes vs.~1156.2 minutes).
Different from the local algorithms, \lcf and \lcp are global cofolding algorithms without restriction on base pair length. 
Surprisingly, \lcf and \lcp's predictions have higher PPV and sensitivity
of intermolecular base pairs.
Furthermore, we apply \lcf to predict the 
RNA-RNA interaction between SARS-CoV-2 gRNA and human U4 snRNA, which has been experimentally studied,
and observe that \lcf's prediction correlates better to the wet lab results. 

\end{abstract}

\dates{}

\begin{document}

\verticaladjustment{-2pt}

\maketitle
\thispagestyle{firststyle}
\ifthenelse{\boolean{shortarticle}}{\ifthenelse{\boolean{singlecolumn}}{\abscontentformatted}{\abscontent}}{}


\vspace{-.5cm}
\label{sec:intro}

\section{Introduction}

\begin{figure}[!t]
\hspace{.2cm}\includegraphics[width=.49\textwidth]{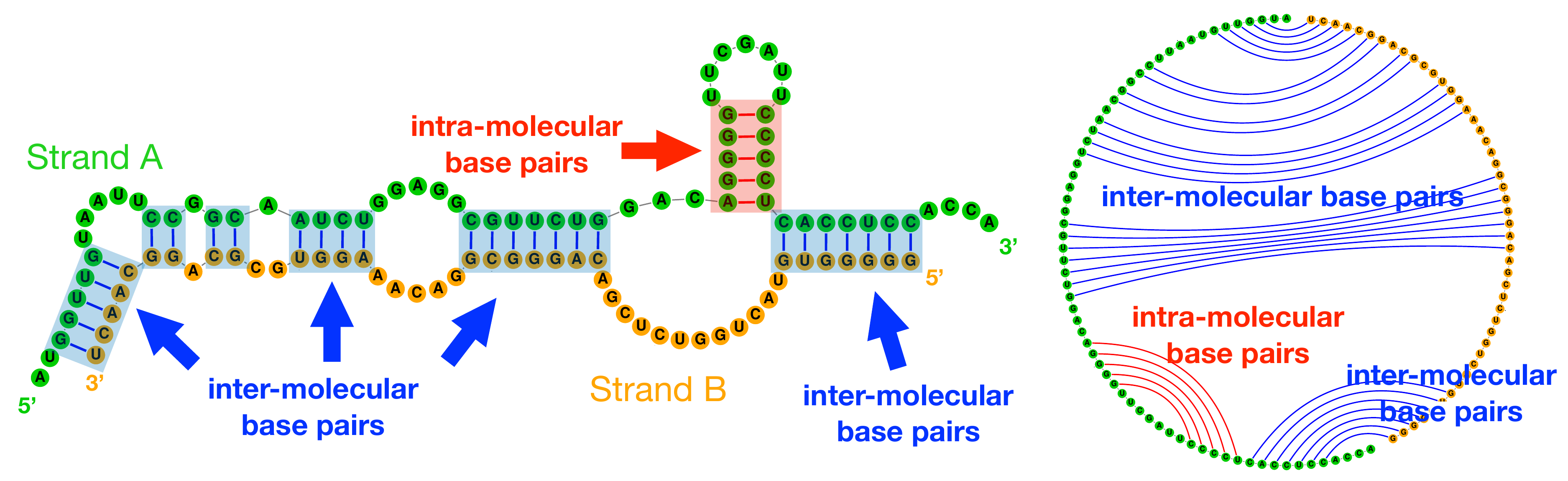} \\[-2.5cm]
\begin{tabular}{cc}
\hspace{-.2cm}\panel{A} & \hspace{5.2cm}\panel{B} \\[2.2cm]
\hspace{-.2cm}\panel{C} &  \\[-.4cm]
\end{tabular}
\resizebox{.49\textwidth}{!}{
\hspace{.3cm}\begin{tabular}{c|c}
{\bf RNA-RNA interaction} & {\bf function} \\
\hline
siRNA-mRNA & mRNA degradation\\
\hline
\multirow{2}{*}{miRNA-mRNA} & mRNA cleavage, destabilization  \\
&and down-regulation \\
\hline
sRNA-mRNA & mRNA silencing \\ 
\hline
gRNA-mRNA & mRNA editing \\
\hline
snRNA-mRNA & RNA splicing and regulation \\
\hline
snoRNA-rRNA & rRNA modification \\
\hline
\multirow{2}{*}{split tmRNA} & rescue of stalled ribosomes; \\
& degradation of defective mRNA
\end{tabular}}
\caption{Two RNA strands can form RNA-RNA complexes through inter-molecular base pairs.
These interacting RNA molecules are widely distributed in nature,
and are involved in multiple biological processes.  
{\bf A}: The secondary structure of the split tmRNA from \daromatica;
two strands are in green and orange, respectively.
The intra-molecular base pairs are in red,
and inter-molecular ones are in blue.
{\bf B}: The corresponding circular plot of structure in {\bf A}.
{\bf C}: Some known RNA-RNA interactions and their functions.
	\label{fig:overview}
}
\end{figure}

RNA strands can interact via inter-molecular base pairing and form RNA-RNA complexes.
In nature, many non-coding RNAs (ncRNAs) function through these RNA-RNA interactions (Fig.~\ref{fig:overview}). 
For instance, 
it is well-known that microRNA (miRNA)  
binds with messenger RNA (mRNA) to mediate mRNA destabilization~\cite{tat2016cotranslational} and cleavage~\cite{xu2016microrna}.
Some longer ncRNAs, 
such as small RNA (sRNA), small nuclear RNA (snRNA) and small nucleolar RNA (snoRNA),
involve in RNA-RNA interactions for splicing regulation~\cite{rogers1980mechanism,mckeown1993role}
and chemical modifications~\cite{kiss2002small}.
A small clade of tmRNAs have a two-piece form (i.e., split tmRNA) and form complexes via inter-molecular base pairs (see Fig.~\ref{fig:overview}A and B).
On the other hand, human designed RNAs that bind specifically to the target RNAs are used for diagnostics
and treatments.
Therapeutic small interfering RNA (siRNA)
triggers RNA interference (RNAi) through siRNA-mRNA interaction~\cite{elbashir2001duplexes,yuan2019approval,hu2020therapeutic};
antisense oligonucleotide (ASO) binds to target RNA to suppress unwanted gene expression or to regulate splicing~\cite{stephenson1978inhibition,dias2002antisense,rinaldi2018antisense};
CRISPR/Cas-13 guide RNA (gRNA) induces specific RNA editing by initially binding to the target region~\cite{wiedenheft2012rna,zhang2018structural,bandaru2020structure}.
Fast and reliable secondary structure prediction of interacting RNA molecules is desired
to further understand these biological processes and
better design diagnostic and therapeutic RNA drugs. 


\begin{table*}[!t]
\centering
\begin{tabular}{ccccccc}
\multirow{2}{*}{\bf system} & {\bf input} & \multirow{2}{*}{\bf output} & {\bf MFE} or    & {\bf base pair} & \multirow{2}{*}{\bf runtime} & {\bf memory} \\
						& {\bf strand(s)} &                     & {\bf partition} & {\bf type}   &                          & {\bf usage} \\[.1cm]
\toprule \\[-.2cm]
{\bf RNAsubopt}~\cite{lorenz+:2011}	&	\multirow{2}{*}{one} 	&	sampled structures	&	\multirow{2}{*}{partition}	&	\multirow{2}{*}{intramolecular}	&	\multirow{2}{*}{$O(n^3)$}	&	\multirow{2}{*}{$O(n^2)$} \\
{\bf RNAplfold}~\cite{bernhart2011rna}	&	 	&	accessibility	&	&		&		&	 \\[.1cm]
\hline \\[-.2cm]
{\bf OligoWalk}~\cite{mathews1999predicting}	&	two	&	binding affinity \& structure	&	both	&	intermolecular	&	$O((n+m)^2)$		&	$O((n+m)^2)$	\\
\bottomrule \\[-.6cm]
{\bf RNAhybrid}~\cite{rehmsmeier+:2004}	&	\multirow{2}{*}{two} 	&	\multirow{2}{*}{binding structure}	&	\multirow{2}{*}{MFE}	&	\multirow{2}{*}{intermolecular}	&	\multirow{2}{*}{$O(nm)$}	&	\multirow{2}{*}{$O(nm)$} \\
{\bf RNAplex}~\cite{tafer+hofacker:2008}	    &	 	&		&		&		&	&	\\[.1cm]
\hline \\[-.2cm]
\multirow{2}{*}{\bf RNAup}~\cite{muckstein2006thermodynamics}	    &	one  	&	accessibility 	                       &	\multirow{2}{*}{partition}	&	intramolecular	&	$O(n^3 w)$	&	$O(n^2)$\\
            &                   two     &   binding affinity \& structure          &                                & both       & $O(n^3 w)+O(n w^5)$               & $O(n^2)+O(n w^3)$   \\[.1cm]
\hline \\[-.2cm]
\multirow{3}{*}{\bf PairFold}~\cite{andronescu2005secondary}	&	one 	&	\multirow{3}{*}{full structure}	&	\multirow{3}{*}{MFE}	&	intramolecular	&	$O(n^3)$	&	$O(n^2)$ \\
                        	&	two 	&	                                &	                     	&	both	&	$O((n+m)^3)$	&	$O((n+m)^2)$ \\
                        	&	multiple 	&	                                &	                     	&	both	&	$O((\sum_i n_i)^3)$	&	$O((\sum_i{n_i})^2)$ \\[.1cm]
\hdashline \\[-.3cm]
{\bf bifold}~\cite{mathews1999predicting}	&	\multirow{2}{*}{two} 	&	\multirow{2}{*}{full structure}	&	MFE	&	\multirow{2}{*}{both}	&	\multirow{2}{*}{$O((n+m)^3)$}	& \multirow{2}{*}{$O((n+m)^2)$}	\\
{\bf RNAcofold}~\cite{bernhart+:2006}	&		&		&	both	&		&		&	\\
\hdashline \\[-.3cm]
{\bf DuplexFold}~\cite{Przybylska+:2009}	&	two	&	binding structure	&	MFE	&	intermolecular	&	$O(n+m)$		&	$O((n+m)^2)$	\\[.1cm]
\bottomrule \\[-.6cm]
{\bf \lcf}	&	\multirow{2}{*}{two} 	&	\multirow{2}{*}{full structure}	&	MFE	&	\multirow{2}{*}{both}	&	\multirow{2}{*}{$O(n+m)$}	& $O(b{\rm log}b(n+m))$	\\
{\bf \lcp}	&		&		&	partition	&		&		& $O(b^2(n+m))$		\\
\end{tabular}
\vspace{.2cm}
\caption{
An overview of existing RNA-RNA interaction prediction tools and our algorithms.
In the runtime and memory usage columns, we denote $n$ and $m$ as the lengths of two sequences, 
$w$ as the binding window size, and $b$ as the beam size in our \lcf and \lcp.
Note that $w$ and $b$ are constants; by default, $w$ is 25 in RNAup, and $b$ is 100 in our algorithms.
\pairfold is a tool that can do multiple sequence folding, so we denote $n_i$ as the length of the $i$th sequence for its multifolding mode.
Our \lcf and \lcp are the only ones that achieve linear runtime with considering both inter- and intramolecular base pairs.
\label{tab:overall}}
\vspace{-.2cm}
\end{table*}

Some existing algorithms and systems are used for predicting RNA-RNA interaction (see Tab.~\ref{tab:overall}).
The stochastic sampling algorithms~\cite{ding+lawrence:2003} and tools, such as Vienna RNAsubopt~\cite{lorenz+:2011}, can be used to calculate the accessibilities
by counting how many of the structures have the region of interest completely unpaired,
where accessibility is an indicator represents if the corresponding region is open for binding. 
The tool OligoWalk calculates the accessibility for binding of complementary oligonuleotides considering either lowest free energy structures or the full folding ensemble~\cite{mathews1999predicting,lu2008efficient}.
Instead of obtaining accessibility from samples, 
Bernhart et al.~\cite{bernhart2011rna} introduced a cubic runtime algorithm to precisely 
compute accessibility.
Widely used as they are, 
however, these methods are designed for analyzing the accessibility property of the target sequence, 
but are not able 
to predict the binding structure given a specific oligo. 

RNAhybrid~\cite{rehmsmeier+:2004} and RNAplex~\cite{tafer+hofacker:2008}
are another group of algorithms for predicting the hybridization sites in a target RNA that interact with small oligos, 
especially for microRNAs,
by scanning along the target RNA and calculating the intermolecular hybridization.
Though being fast,
they are less informative and less accurate due to omitting the competing intermolecular and intramolecular base pairs~\cite{lai2016comprehensive,umu+gardner:2017}. 
To address this, accessibility-based method is proposed.
As an example,
RNAup~\cite{muckstein2006thermodynamics}
firstly calculates the accessibility of windows of interest,
then computes the binding energy reward of each window for a given oligo,
and finally combines the target region's accessibility and binding reward together to obtain binding affinity.
The drawback of RNAup (as well as other accessibility-based tools) is the slowness:
its first step, accessibility computation for multiple windows, 
employs a $O(n^3w)$ algorithm, where $n$ is the target sequence length and $w$ is the window size,
resulting in a substantially slow down compared to RNAhybrid and RNAplex.

Aiming to compute the binding affinity and predict the binding region,
RNAhybrid, RNAplex and RNAup are not able to predict the complete binding conformation of two sequences.
However, the {\em joint structure} consisting of both the intramolecular base pairs and intermolecular base pairs is desired in many cases.
Fig.~\ref{fig:overview}A and B illustrate the secondary structure in the region of interaction
of the split tmRNA from {\textit {D.~aromatica}}~\cite{DiChiacchio+:2016},
showing that both intramolecular and intermolecular base pairs exist in the binding region.
To predict the joint structure, 
several tools, such as bifold~\cite{mathews1999predicting}, \pairfold~\cite{andronescu2005secondary}, \viennarnacofold~\cite{bernhart+:2006c} and NUPACK~\cite{Dirks+:2007}, were developed.
The basic framework of these tools are to concatenate two input sequences as a single sequence,
and predict the whole secondary structure of the concatenated sequence
based on the classical dynamic programming algorithms. 
With some differences in implementation, 
the runtime of these algorithms are all $O((n+m)^3)$, 
where $n$ and $m$ are the lengths of the two strands,
preventing them to be applied to long sequences,
for instance, long mRNAs and some full-length viral genomes.

To accelerate and scale up the prediction of the joint structure
we propose \lcf and \lcp, which follow the ``concatenation'' strategy to simplify two-strand cofolding into classical single-strand folding,
and predict both intramolecular and intermolecular interactions.
Different from previous cubic runtime algorithms, \lcf and \lcp adopt a left-to-right dynamic programming and further apply beam pruning heuristics to reduce its runtime to linear-time.
Specifically, \lcf predicts the minimum free energy structure of two strands,
while \lcp computes partition function and base pairing probabilities,
and can output assembled structures with downstream algorithms such as MEA~\cite{do+:2006} and ThreshKnot~\cite{Zhang+:2019}.
Unlike other {\it local} cofolding algorithms, 
\lcf and \lcp are {\it global} linear-time algorithms, 
i.e., they do not impose any limitations on base pairing distance.

We compare the efficiency and scalability of our algorithms to \viennarnacofold.
and confirm that the runtime and memory usage of \lcf and \lcp scale linearly against combined sequence,
while \rnacofold scales cubically in runtime and quadratically in memory usage.
\lcf and \lcp are orders of magnitude faster than \rnacofold.
On the longest data point in the benchmark dataset that \rnacofold can run (26,190~\nts),
\lcf is $86.8 \times$ faster than \rnacofold MFE mode,
and \lcp is $642.3 \times$ faster than \rnacofold partition function mode.
Notably, \rnacofold cannot finish any sequences longer than 32,767~\nts,
but our \lcf and \lcp have no limitation of sequence length internally, 
and can scale up to sequences of length 100,000~\nts in 2.2 and 6.9 minutes, respectively.
With respect to accuracy, 
\lcf and \lcp's predictions are more accurate with respect to Sensitivity (the fraction of known pairs correctly predicted) and Positive Predictive Value (PPV; the fraction of predicted pairs that are in the accepted structure).
Compared with \rnacofold MFE, the overall PPV and Sensitivity of \lcf increase +4.0\% and +11.6\%, respectively;
compared with \rnacofold MEA, \lcp MEA gains improvement of +2.9\% on PPV and +5.7\% on sensitivity;
compared with \rnacofold TheshKnot, \lcp TheshKnot increases +2.4\% and +5.5\% on PPV and sensitivity, respectively.
Furthermore, 
we demonstrate that our predicted interaction correlates better to the wet lab results
of the RNA-RNA interaction between SARS-CoV-2 gRNA and human U4 snRNA,
showing that our algorithms can be used as a fast and reliable computational tool in the genome studies.

\label{sec:algorithm}

\section{Algorithms}


\begin{figure*}[!thb]
\begin{center}
\includegraphics[width=.99\textwidth]{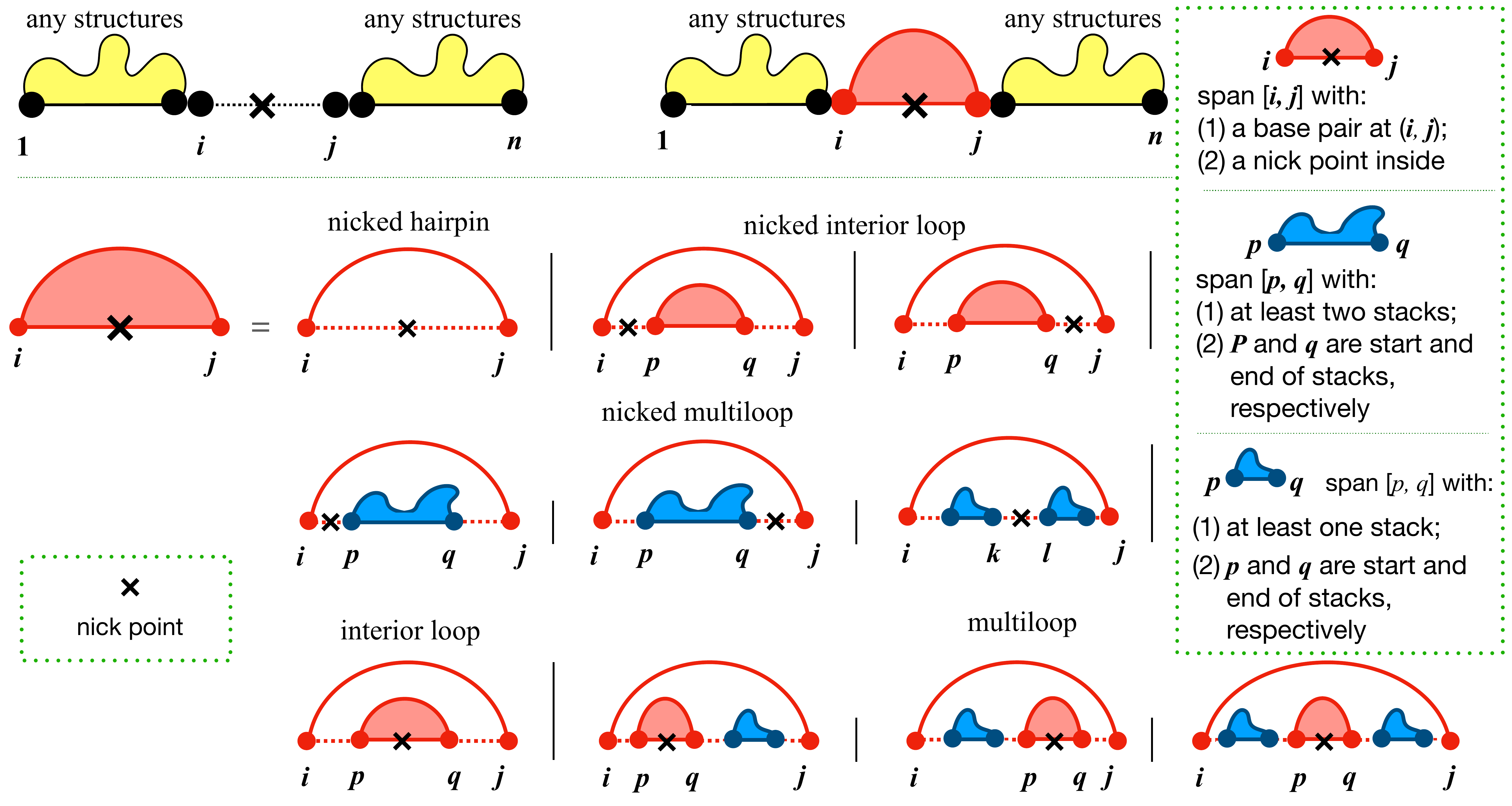} \\[-9.3cm]
\begin{tabular}{cc}
\hspace{-17.cm}\panel{A} & \hspace{-2.5cm}\panel{B} \\[2.2cm]
\hspace{-17.cm}\panel{C} &  \\
\end{tabular}
\vspace{6.2cm}
\caption[]{The relative positions of the nick point when concatenating two strands for zuker-style cofolding.
{\bf A}: The nick point is not covered by a base pair, i.e., there is no intermolecular base pairs.
{\bf B}: The nick point is covered by an intermolecular base pair;
note that only in this case two stands form a RNA-RNA complex.
{\bf C}: The breakdown cases of the interacting span $[i, j]$ in {\bf B}.
When the nick point is directly covered by the outside intermolecular base pair $(i, j)$ (the first and second rows in {\bf C}),
they form no more hairpins, interior loops or multiloops, but exterior loops,
so we call them the corresponding ``nicked'' loops.
But when the nick point is covered by a nested base pair $(p, q)$,
they are still normal interior loops and multiloops (the third row in C).
\label{fig:fake_loops}}
\end{center}
\vspace{-0.3cm}
\end{figure*}
\subsection{Extend Single-strand Folding to Double-strand Folding by concatenation}

Both \linearcofold and \linearcopartition take two RNA sequences as input, 
and simplify the two-strand cofolding to the single-strand folding via concatenating two input RNAs. 
Formally, we denote the two RNA sequences as $\vecxa=x_1^a x_2^a...x_n^a$ and $\vecxb=x_1^b x_2^b...x_m^b$, 
where $n$ and $m$ are the lengths of \vecxa and \vecxb, respectively.
Thus, the new concatenated sequence of length $n+m$
can be denoted as $\vecx=x_1 x_2...x_n x_{n+1} x_{n+2}...x_{n+m}$, 
where the nick point is between nucleotides $x_n$ and $x_{n+1}$.

After this transformation, 
the classical dynamic programming algorithm for single-strand folding~\cite{nussinov+jacobson:1980,zuker+stiegler:1981} can be applied to the concatenated sequence.
One thermodynamic change needs to be considered for this extension is 
that a structure that contains intermolecular base pairs incurs a stability penalty for intermolecular initiation~\cite{xia+:1998}.
Formally, in the Nussinov system,
we denote the free energy change of the first intermolecular base pair $(i, j)$ as $\zeta(\vecx, i, j)$,
which differentiates it from that of the normal base pair $(p, q)$, $\xi(\vecx, p, q)$.
Note that $(i,j)$ is the innermost base pair that contains the nick point,
while other intermolecular base pairs do not incur an addition stability cost.
Besides, the free energy change of the unpaired base $k$ is denoted as $\delta(\vecx, k)$.
Thus, the free energy change $\Delta G^{\circ}(\vecx, \vecy)$ of the concatenated sequence $\vecx$ and its structure $\vecy$ 
($(i,j) \in \vecy$) 
can be decomposed as: 
\begin{equation}
\label{equ:decompose}
\Delta G^{\circ}(\vecx, \vecy) = \! \! \! \! \! \!  \! \!  \sum_{k \in {\rm unpaired}(\vecy)}\! \! \! \! \! \!  \! \! {\delta(\vecx, k)} + {\zeta(\vecx, i, j)} + \! \! \! \!  \! \! \!  \! \sum_{\substack{(p,q) \in {\rm pairs}(\vecy) \\ (p,q) \neq (i,j)}}\! \! \! \! \!  \! \! \! {\xi(\vecx, p, q)} \\
\end{equation}
Note that if there is no base pair closing the nick point, 
i.e., the two strands do not interact with each other, 
two-strand cofolding is simply single-strand folding of two strands separately. 

Next, we consider the Zuker system based on the Turner energy model~\cite{Zuker+Sankoff:1984,Mathews+:1999,Mathews+:2004}.
More sophisticated than the Nussinov model, 
the Zuker and Turner's scoring system is based on four types of loops: exterior loops, hairpin loops, interior loops (where a bulge loop with unpaired nucleotides only on one side is considered a type of interior loop) and multiloops.
In Fig.~\ref{fig:fake_loops}, we illustrate the relative positions of the nick point in these four types of loops.
For the external loop, the nick point can be either covered by a base pair or not (Fig.~\ref{fig:fake_loops}A and B).
If an intermolecular base pair $(i,j)$ closing the nick point, the span $[i,j]$ can be further decomposed into nicked hairpin, nicked interior loop and nicked multiloop (Fig.~\ref{fig:fake_loops}C) based on the type of loops it enclosed. 
Specifically, the nicked hairpin loop only requires $i \leq n < j$,
while the nicked interior loop has an inner loop from position $p$ to $q$,
and requires either $i \leq n < p$ or $q \leq n < j$;
see the first row of Fig.~\ref{fig:fake_loops}C for an illustration.
The nicked multiloop is more complicated (the second row of Fig.~\ref{fig:fake_loops}C):
\begin{itemize}
\item the nick point is at the leftmost unpaired region, i.e., it is between $i$ and $p$ where $p$ is the 5' end of the first multibranch;
\item the nick point is at the rightmost unpaired region, i.e., it is between $q$ and $j$ where $q$ is the 3' end of the last multi-branch;
\item the nick point is in the middle, i.e., it is between $k$ and $l$ which are the 3' end and the 5' end of two consecutive multi-branches, respectively.
\end{itemize}


\begin{figure}[!t]
\hspace{.1cm}\includegraphics[width=.7\textwidth]{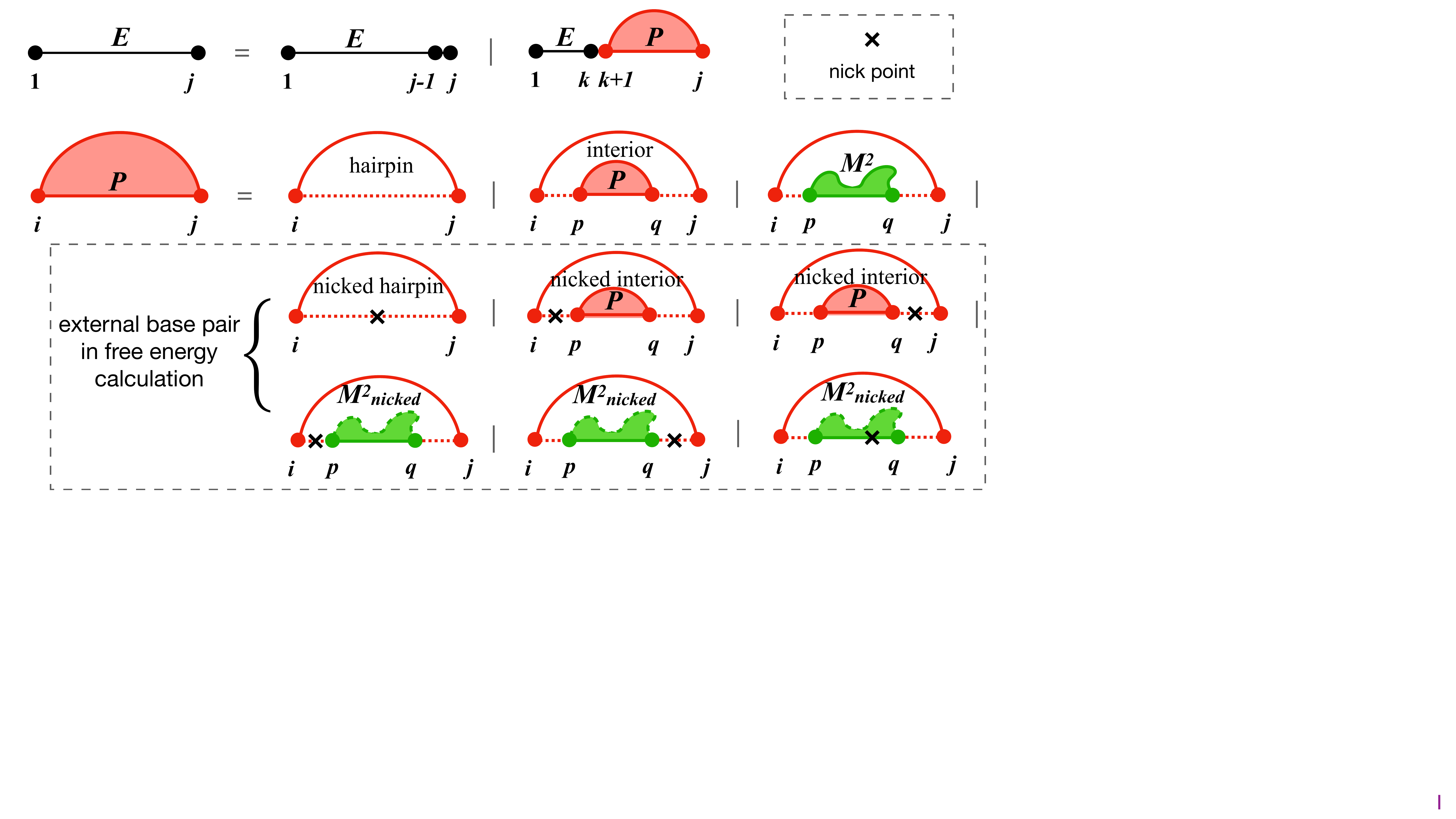} \\[-2.6cm]
\hspace{.1cm}\includegraphics[width=.7\textwidth]{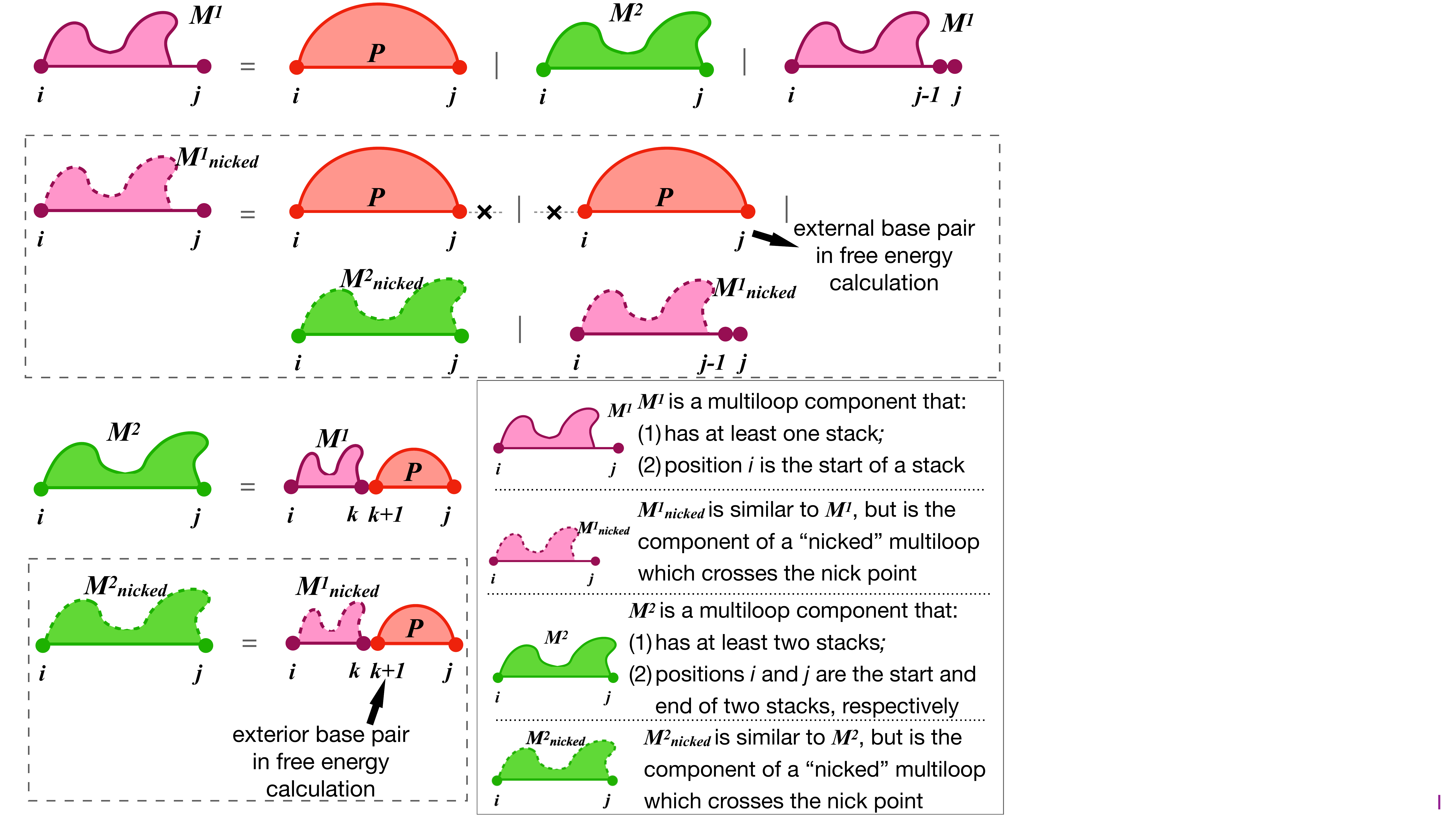} \\[-.5cm]
\caption{Deductive system of \lcf and \lcp based on the Zuker system. 
	\label{fig:deductive}
For single-strand folding (ignoring dashed boxes), four states $\statee$, $\statep$, $\statem$ and $\statemm$ are defined to retain the MFE structure for the span $[i, j]$, 
where $\statep$ requires $i$ paired with $j$,  
$\statem$ and $\statemm$ are the components of multiloops. 
To extend to two-strands cofolding (adding dashed),
first, \lcf takes into consideration the nicked hairpin, nicked interior loop and nicked multiloop for state $\statep$.
Besides, 
\lcf also adds two states $\statemn$ and $\statemmn$ to model the components of nicked multiloops. 
More importantly, 
the innermost base pairs enclosing the nick point to form $\statep$ (first dashed box), 
as well as the closing base pairs of branches of $\statep$ when forming  $\statemn$ and $\statemmn$ (second and third dashed boxes) 
are treated as external base pairs since the nick points are exterior, 
i.e., these base pairs are not closed by any base pairs in each single strand. 
Besides, 
\lcf only picks up the MFE structure,
while \lcp sums up all possible structures for each state. 
}
\end{figure}

Such nicked loops are considered to be exterior loops when calculating their free energy change.
Note that the nick point only affects the innermost loop that directly covers it;
the loops are still normal interior loops and multiloops 
in the case that the nick point is covered by another base pair $(p, q)$ where $i<p<q<j$,
shown in the third row of Fig.~\ref{fig:fake_loops}C.
In addition, we add the intermolecular initiation free energy cost for dimers.

\subsection{\lcf Algorithm}
\label{alg:lcf_alg}

 \lcf aims to predict the minimum free energy (MFE) structure of double-strand RNAs in linear runtime without imposing a limit on base pair length.
Formally, \lcf finds the MFE structure $\vecyhat$ 
among all possible structures $\mathcal{Y}(\vecx)$ under the given energy model $\vecw$: 
\begin{equation*}
\vecyhat = \argmin_{\vecy \in \mathcal{Y}(\vecx)}{\Delta G^{\circ}_{\vecw}(\vecx, \vecy)}
\end{equation*}

Inspired by \linearfold~\cite{huang+:2019},
\lcf adopts a left-to-right dynamic programming (DP), 
in which we scan and fold the combined sequence from left to right. 
Fig.~\ref{fig:algorithm_lcf} presents the pseudocode of \lcf based on the revised Nussinov-Jacobson energy model.
In the pseudocode, we use a hash table $\Cf{i}{j}$ to memorize the best score for each span $[i,j]$.
At each step $j$, 
two actions, SKIP (line~\ref{line:skip_lcf}) and POP (line~\ref{line:pop_fake_lcf} and~\ref{line:pop_normal_lcf}), are performed,
where SKIP extends $\Cf{i}{j-1}$ to $\Cf{i}{j}$ by adding an unpaired base $\vecy_j$ =``$\cdot$'' to the right of the best substructure on the span $[i,j-1]$, 
and POP combines $\Cf{i}{j-1}$ with an upstream span $\Cf{k}{i-2}$ ($k < i$) and updates the resulting $\Cf{k}{j}$ if $\vecx_{i-1}$ can be paired with $\vecx_{j}$.
Note that this new DP algorithm is equivalent to the classical algorithm in the sense that they both find the MFE structure in cubic time,
however, such left-to-right fashion allows applying beam pruning,
which retains the top $b$ states with lower folding free energy change at each step $j$ (line~\ref{line:beamprune}).
As a result, the time complexity of \lcf is $O(nb^2)$, where $b$ is the beam size.
It is clear in the pseudocode that \lcf does not impose any constraints on base-pairing distance,
which is different from the local folding approximation.
To extend to two-strands cofolding,  
\lcf distinguishes between intramolecular and intermolecular base pairs following Equation~\ref{equ:decompose}, 
and rewards them with different energy scores (from line~\ref{line:start} to line~\ref{line:end}). 

Compared to the Nussinov-Jacobson energy model, 
the Zuker system based on the Turner energy mode defines more states to represent different types of loops. 
Formally, for single-strand folding, 
state $\statee$, $\statep$, $\statem$ and $\statemm$ retain the MFE structure for the span $[i, j]$, 
where $\statep$ requires $i$ paired with $j$,  
$\statem$ has at least one branch with  $i$ as the 5' end of the leftmost branch, 
and $\statemm$ contains at least two branches with $i$ and $j$ as the 5' end and the 3' end of the leftmost and rightmost branches, respectively (Fig.~\ref{fig:deductive} except for dashed boxes). 
$\statem$ and $\statemm$ are the components of multiloops. 
Extending to two-strand cofolding (dashed boxes in Fig.~\ref{fig:deductive}),
\lcf takes into consideration the nicked hairpin, nicked interior loop and nicked multiloop for state $\statep$. 
In addition,
\lcf also adds two states $\statemn$ and $\statemmn$ to model the components of nicked multiloops. 
Compared to $\statem$ and $\statemm$,
the closing pairs of branches (state $\statep$) in $\statemn$ and $\statemmn$ are scored as an external base pairs since the nick point breaks the multiloop, 
i.e., these closing pairs are not enclosed by any base pairs in each single strand. 
Similarly, the innermost base pair enclosing the nick point is also scored as an external base pair (dashed boxes for state $\statep$).  
Besides, the intermolecular initiation free energy is added to the innermost base pair across the nick point in the Zuker system.

\subsection{\lcp Algorithm}


Beyond the MFE structure,
a partition function and base-pairing probabilities of cofolding two RNA strands, 
and their assembled structure from the ensemble (e.g., MEA structure) are desired in many cases.
A partition function $Q(\vecx)$ sums the equilibrium constants of all possible secondary structures in the ensemble. 
Using the revised Nussinov-Jacobson energy model defined in Sec.~\ref{alg:lcf_alg},
the partition function of two interacting RNAs can be formalized as:
\begin{equation*}
\begin{split}
	Q(\vecx) &= \sum_{\vecy \in \mathcal{Y(\vecx)}}  e^{-\frac{\Delta G^{\circ}_{\vecw}({\vecx, \vecy})}{RT}} \\
			&= \sum_{\vecy \in \mathcal{Y(\vecx)^{\prime}}} (\prod_{k \in {\textrm {unpaired}}(\vecy)} \! \! \! \! \! \! \! \! \! e^{-\frac{\delta(\vecx, k)}{RT}})  \cdot
			e^{-\frac{\zeta(\vecx, i, j)}{RT}}  \cdot
			( \! \! \! \! \!  \! \! \! \! \! \prod_{\substack{(p,q) \in {\rm pairs}(\vecy) \\ (p,q) \neq (i,j)}}  \! \! \! \! \! \! \!  \! \! e^{-\frac{\xi(\vecx, p, q)}{RT}} ) \\
			&+ \sum_{\vecy \in \mathcal{Y(\vecx)^{\prime\prime}}} (\prod_{k \in {\textrm {unpaired}}(\vecy)} \! \! \! \! \! \! \! \! \! e^{-\frac{\delta(\vecx, k)}{RT}})  \cdot
			( \! \! \! \! \!  \! \! \! \! \! \prod_{\substack{(p,q) \in {\rm pairs}(\vecy) }}  \! \! \! \! \! \! \!  \! \! e^{-\frac{\xi(\vecx, p, q)}{RT}} )
\end{split}
\end{equation*}
where $\mathcal{Y(\vecx)^{\prime}}$ is the set of structures, 
in which interactions exist between two strands, 
while $\mathcal{Y(\vecx)^{\prime\prime}}$ enumerates the rest of structures of $\mathcal{Y(\vecx)}$, 
in which two strands do not interact with each other, 
and therefore no special treatment is needed for the nicked base pair $(i, j)$. 
Additionally, $R$ is the universal gas constant and $T$ is the absolute temperature.

We further extend \lcf to \lcp based on the inside-outside algorithms following \linearpartition~\cite{zhang2020linearpartition}, 
which calculates the local partition function $\Qf{i}{j}$ in a left-to-right order. 
Fig.~\ref{fig:algorithm_lcp} shows a simplified pseudocode based on the Nussinov-Jacobson model.
\lcp consists of two major steps: partition function calculation (``inside phase'') and base-pairing probability calculation (``outside phase''), which is symmetrical to the inside phase but in a ``right-to-left'' order.
The inside phase updates a hash table $\Qf{i}{j}$ to keep partition function for each span $[i, j]$, 
and the outside phase maintains another hash table $\Qhatf{i}{j}$ with the ``outside partition function'', 
which represents an ensemble of structures outside the span $[i, j]$. 
Based on $\Qf{i}{j}$, $\Qhatf{i}{j}$ and the partition function for the combined sequence $\Qf{1}{n+m}$, the base-pairing probability $p_{i,\,j}$ can be derived if position $i$ can be paired with $j$ (line~\ref{line:bpp}). 
Similar as \lcf, 
two actions SKIP (line~\ref{line:skip_lp}) and POP are performed, 
and POP action distinguishes intermolecular base pairs from intramolecular pairs and rewards them with different energy parameters (line~\ref{line:pop_fake} and \ref{line:pop_normal}) in both inside and outside phases. 

\label{sec:results}

\section{Results}

\begin{figure*}[!htb]
\begin{tabular}{ccc}
\hspace{-6.8cm}\panel{A} & \hspace{-8.4cm}\panel{B} & \hspace{-8.7cm}\panel{C} \\[-.3cm]
\hspace{-1.5cm}\includegraphics[width=0.48\textwidth]{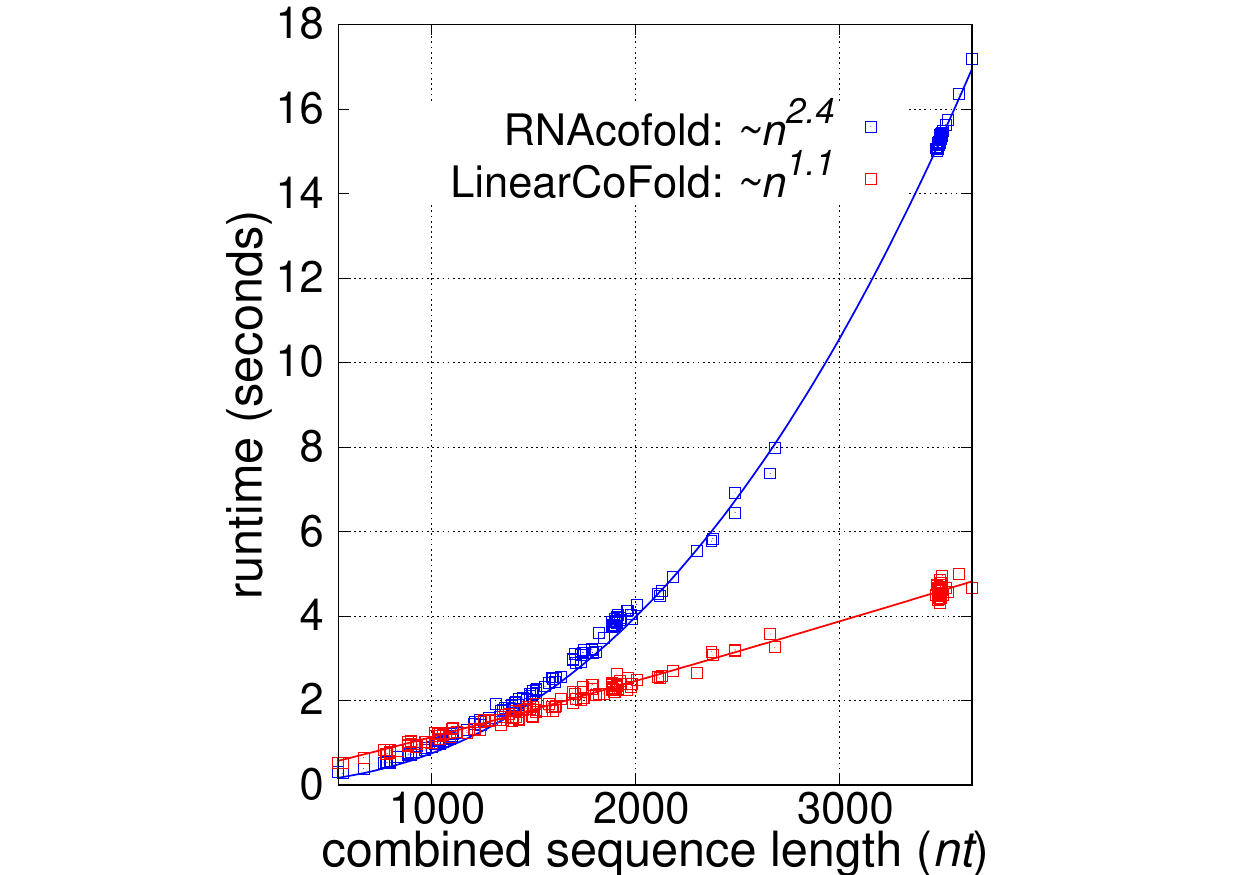} &
\hspace{-3cm}\includegraphics[width=0.48\textwidth]{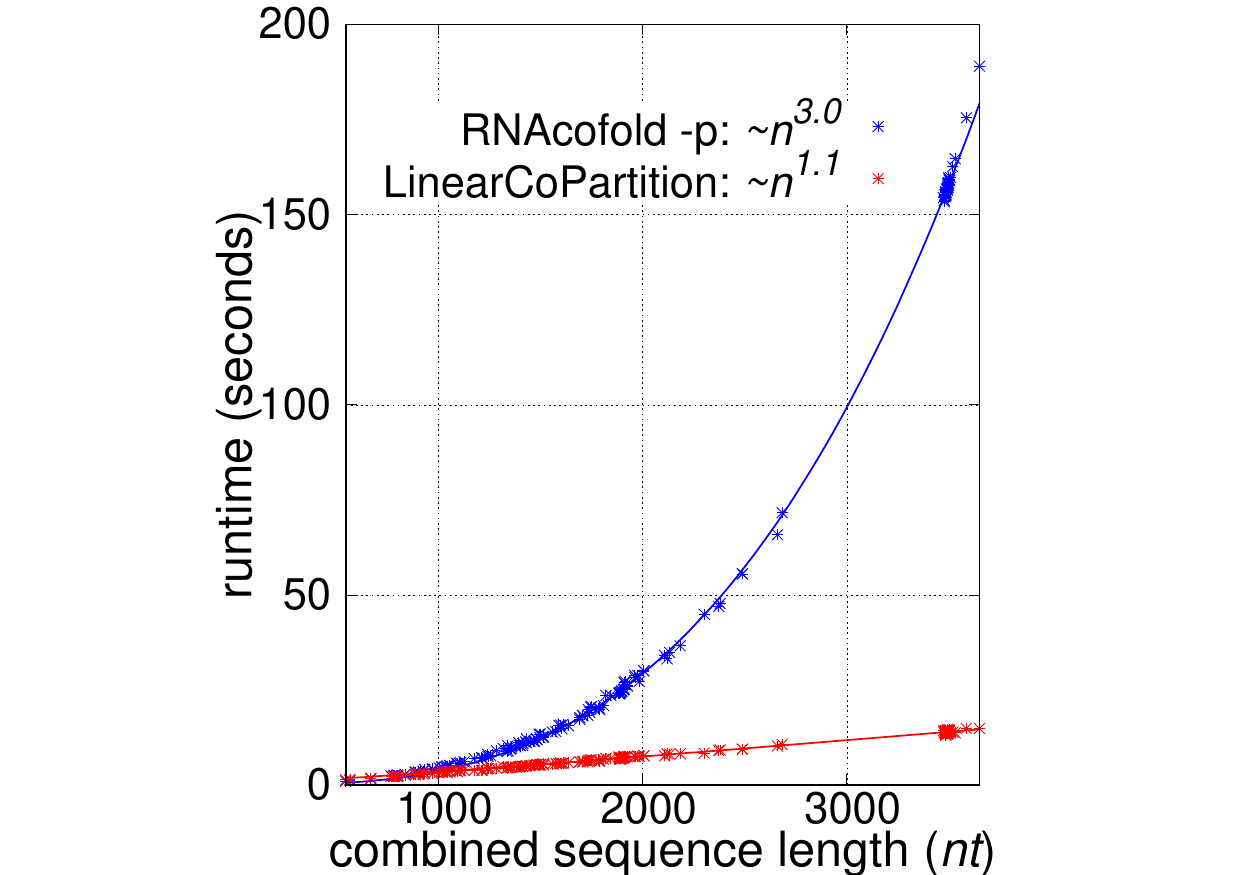} &
\raisebox{.06cm}{\hspace{-2.8cm}\includegraphics[width=0.475\textwidth]{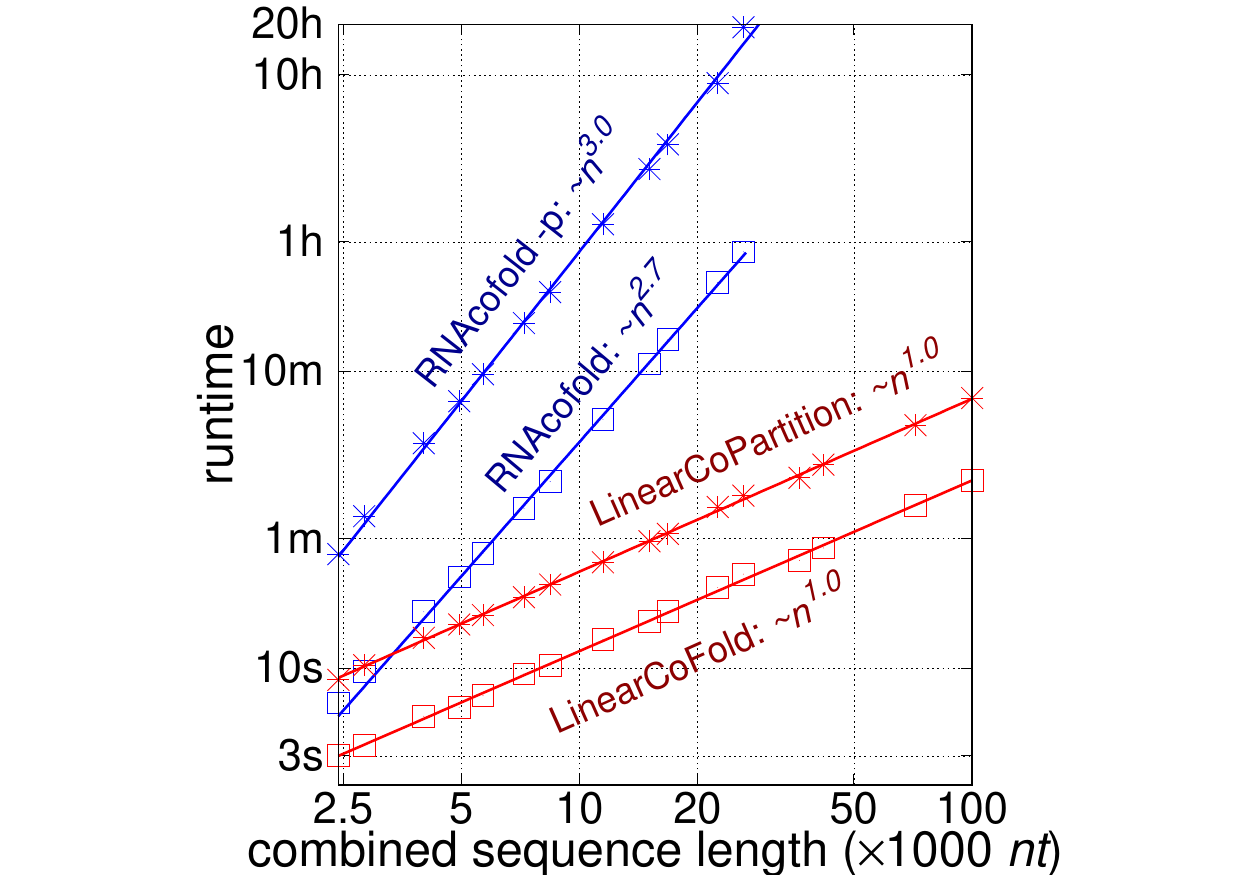}} \\[.2cm]
\end{tabular}
\begin{tabular}{cc}
\hspace{-8.3cm}\panel{D} & \hspace{-8.4cm}\panel{E} \\[-.2cm]
\hspace{-.0cm}\includegraphics[width=0.48\textwidth]{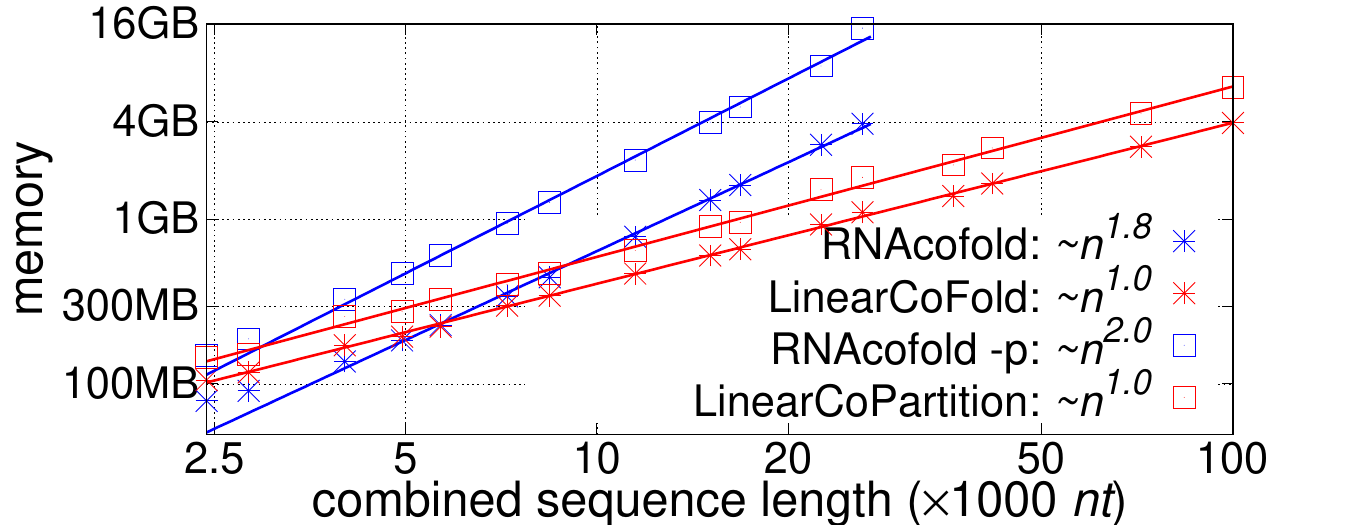} &
\raisebox{1.5cm}{\hspace{-.1cm}\begin{tabular}{c|c|cc} 
{\bf dataset}        & {\bf Meyer Dataset } & \multicolumn{2}{c}{\bf TargetScan Dataset} \\
{\bf example length}       & 255+3,396~\nts & \multicolumn{2}{c}{28+26,162~\nts} \\[0.05cm]
\toprule 
\multirow{2}{*}{\bf metric} & runtime        & runtime & memory-used \\
& (seconds)        & (minutes) & (GB) \\[0.05cm]
\hline 
{\bf \rnacofold} & 17.2 & 52.1 & 3.9\\
{\bf \lcf} & 4.7 & 0.6 & 1.1\\[0.05cm]
\hline 
{\bf \rnacofold -p} & 189.1 & 1156.19 &15.1\\
{\bf \lcp} & 14.9 & 1.8 &1.8\\
\end{tabular}}
\end{tabular}
\caption{Runtime and Memory usage comparisons between \rnacofold and our algorithms.
{\bf A-B}: runtime against sequence length on the Meyer dataset; \rnacofold (MFE mode) and \lcf and compared in {\bf A}, while \rnacofold -p (partition function mode) and \lcp and compared in {\bf B}.
{\bf C}: runtime against sequence length on the TargetScan dataset.
{\bf D}: memory usage against sequence length on the TargetScan dataset. Note that {\bf C} and {\bf D} are plotting in the log-log scale.
{\bf E}: the performance comparisons on two selected examples from the two dataset. The example from the Meyer dataset is one of the sequences that have the longest combined length, and the example from the TargetScan dataset is the longest one that \rnacofold can run.
\label{fig:time_mem}}
\end{figure*}

\subsection{Datasets}

We compared the performance of \lcf and \lcp to \rnacofold on two datasets.
The first dataset, collected by Lai and Meyer~\cite{lai2016comprehensive},
contains 109 pairs of bacterial sRNA-mRNA sequences and 52 pairs of fungal snoRNA-rRNA sequences 
with annotated ground truth of intermolecular base pairs.
The combined sequence length in this dataset
ranges from 546~\nts to 3,651~\nts.
We refer this dataset as the Meyer dataset in the paper.
The second dataset contains 16 miRNA-mRNA pairs from the TargetScan database~\cite{agarwal2015predicting}.
We first sampled 16 mRNA sequences ranging from 2,411 to 100,275~\nts,
and sampled 16 miRNA sequences ranging from 15~\nts to 28~\nts,
and then randomly assemble them into 16 miRNA-mRNA pairs with combined sequence length (i.e., $n+m$) ranging from 2,432 to 100,297~\nts.
We refer this dataset as the TargetScan dataset in the paper.
For benchmark, 
we used a Linux machine (CentOS 7.9.2009) with 2.40 GHz Intel Xeon E5-2630
v3 CPU and 16 GB memory, and gcc 4.8.5.


\subsection{Efficiency and Scalability}

We first investigated the efficiency of \lcf and \lcp by plotting the runtime against the combined sequence length,
and compared them to \viennarnacofold on the Meyer dataset,
whose sequences are relatively shorter  than the TargetScan dataset.
Fig.~\ref{fig:time_mem}A and B clearly shows that our \lcf and \lcp both achieve linear runtime with the combined sequence length;
in contrast, 
\rnacofold runs in nearly cubic time (MFE mode, Fig.~\ref{fig:time_mem}A) or exactly cubic time (partition-function mode, Fig.~\ref{fig:time_mem}B) in practice.
Our algorithms are substantially faster than \rnacofold on long sequences ($n+m> 1,500$~\nts).
For one of the longest combined sequences with length of 3,651 (255+3,396)~\nts, 
\lcf is $3.7 \times$ faster than \rnacofold MFE mode (4.7 vs.~17.2 seconds),
and \lcp is $12.7 \times$ faster than \rnacofold partition-function mode (14.9 vs.~189.1 seconds).


Fig.~\ref{fig:time_mem}C presents the efficiency and scalability comparisons on the TargetScan dataset in log-log scale.
The two blue lines illustrate that \rnacofold's runtime scales (close to) cubically on the long sequences,
and the two red lines confirm that the runtime of \lcf and \lcp are indeed linear.
We also observed that \lcf and \lcp can scale to sequences of length 100,000~\nts in 2.2 and 6.9 minutes, respectively,
while \rnacofold cannot process any sequences with combined sequence length longer than 32,767~\nts.
For the longest sequence pair (combined sequence length 26,190~\nts) in the dataset that \rnacofold can run,
\lcf is $86.8 \times$ faster than \rnacofold MFE mode (0.6 vs.~52.1 minutes),
and surprisingly, \lcp is $642.3 \times$ faster than \rnacofold partition-function mode (1.8 vs.~1156.2 minutes).

The memory usage on the TargetScan dataset is shown in Fig.~\ref{fig:time_mem}D.
From the plots in log-log scale, 
we can see that the memory required by our \lcf and \lcp increases linearly with the sequence length,
while it scales quadratically for \rnacofold.
For the longest one within the scope of \rnacofold, 
\lcf takes 28.2\% of memory compared to \rnacofold MFE mode (1.1 vs.~3.9 GB),
and \lcp takes only 11.9\% of memory compared to \rnacofold partition-function mode (1.8 vs.~15.1 GB).

\subsection{Accuracy}

\newcommand\crule[3][black]{\textcolor{#1}{\rule{#2}{#3}}}
\newcommand\filledcirc[1][black]{{\color{#1}\bullet}\mathllap{\color{#1}\circ}}

\begin{figure*}[!t]
\begin{tabular}{ccc}
\hspace{-6.cm}\panel{A} & \hspace{-6.8cm}\panel{B} & \hspace{-6.5cm}\panel{C} \\[-.4cm]
{\raisebox{.0cm}{\hspace{-.8cm}\includegraphics[width=0.36\textwidth]{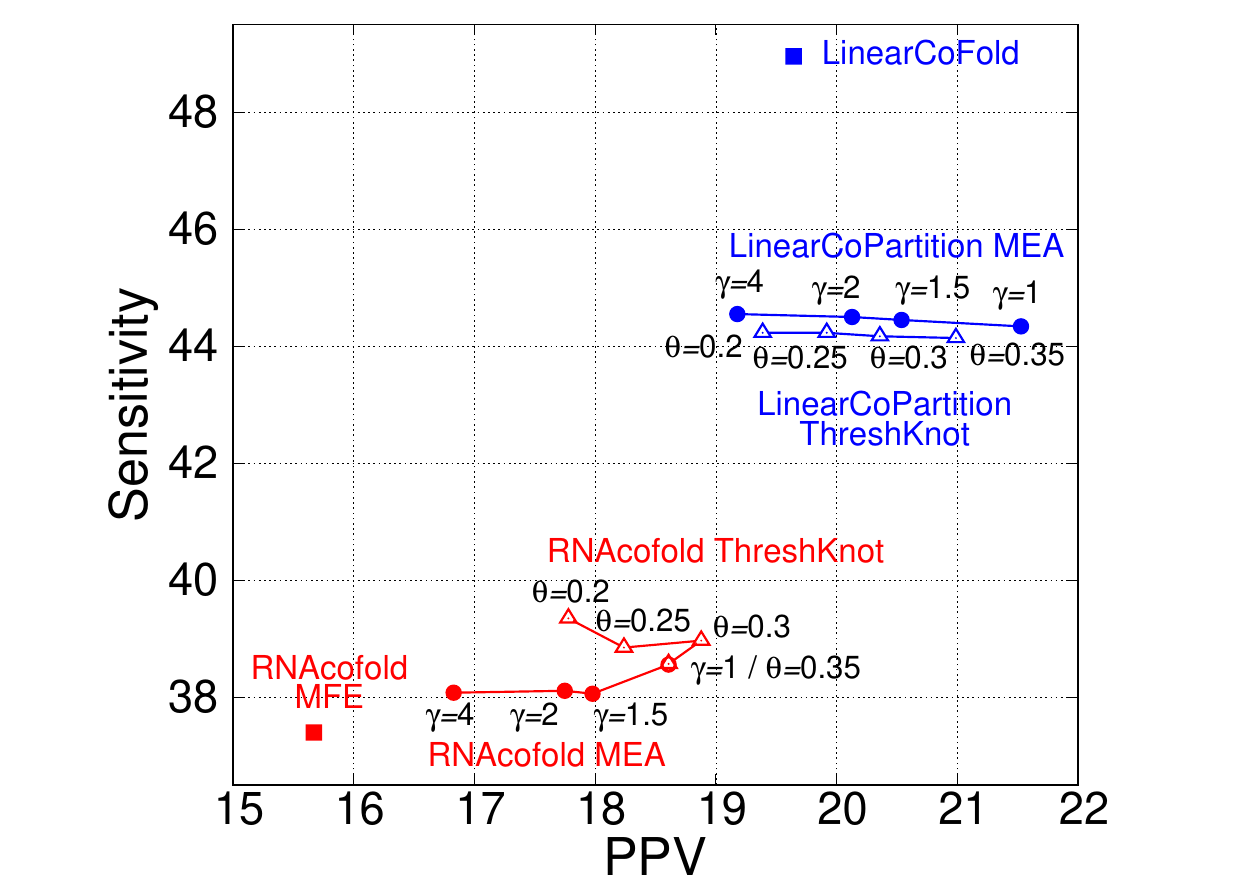}}} & 
{\raisebox{.2cm}{\hspace{-1.15cm}\includegraphics[width=0.373\textwidth]{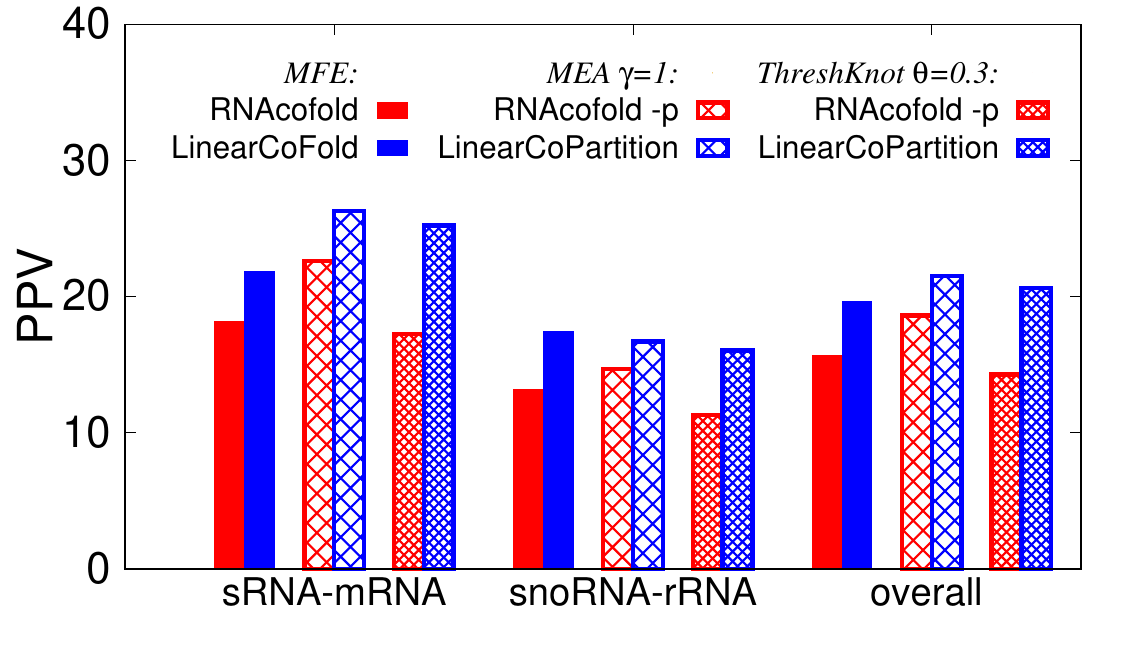}}} &
{\raisebox{.2cm}{\hspace{-.6cm}\includegraphics[width=0.373\textwidth]{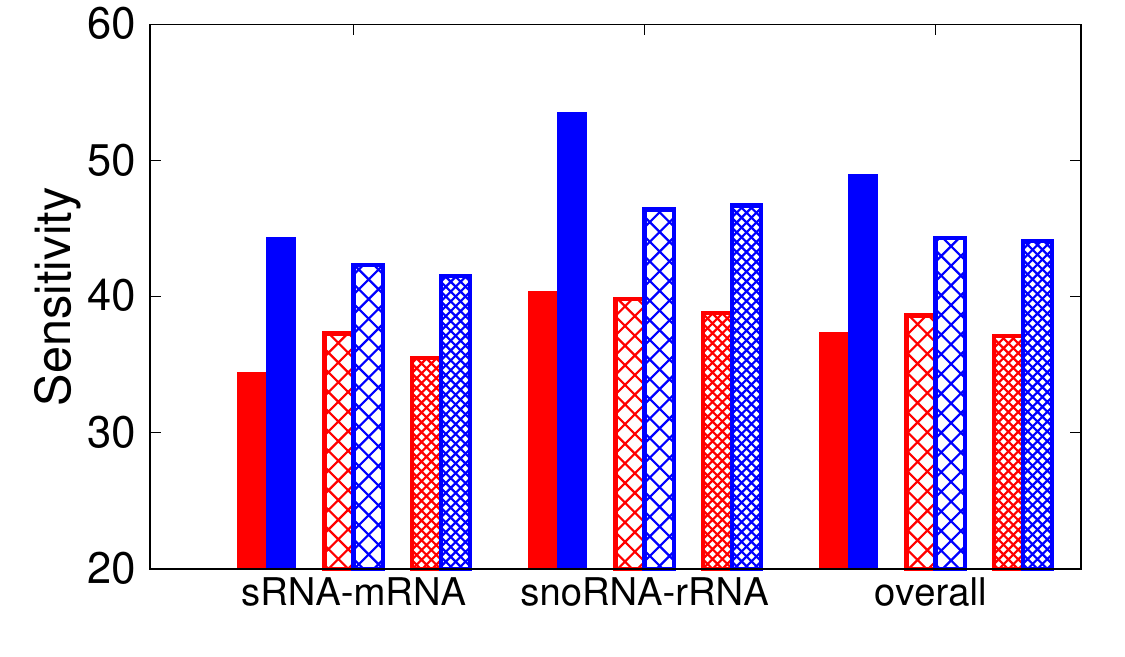}}} \\
\end{tabular}
\includegraphics[width=0.99\textwidth]{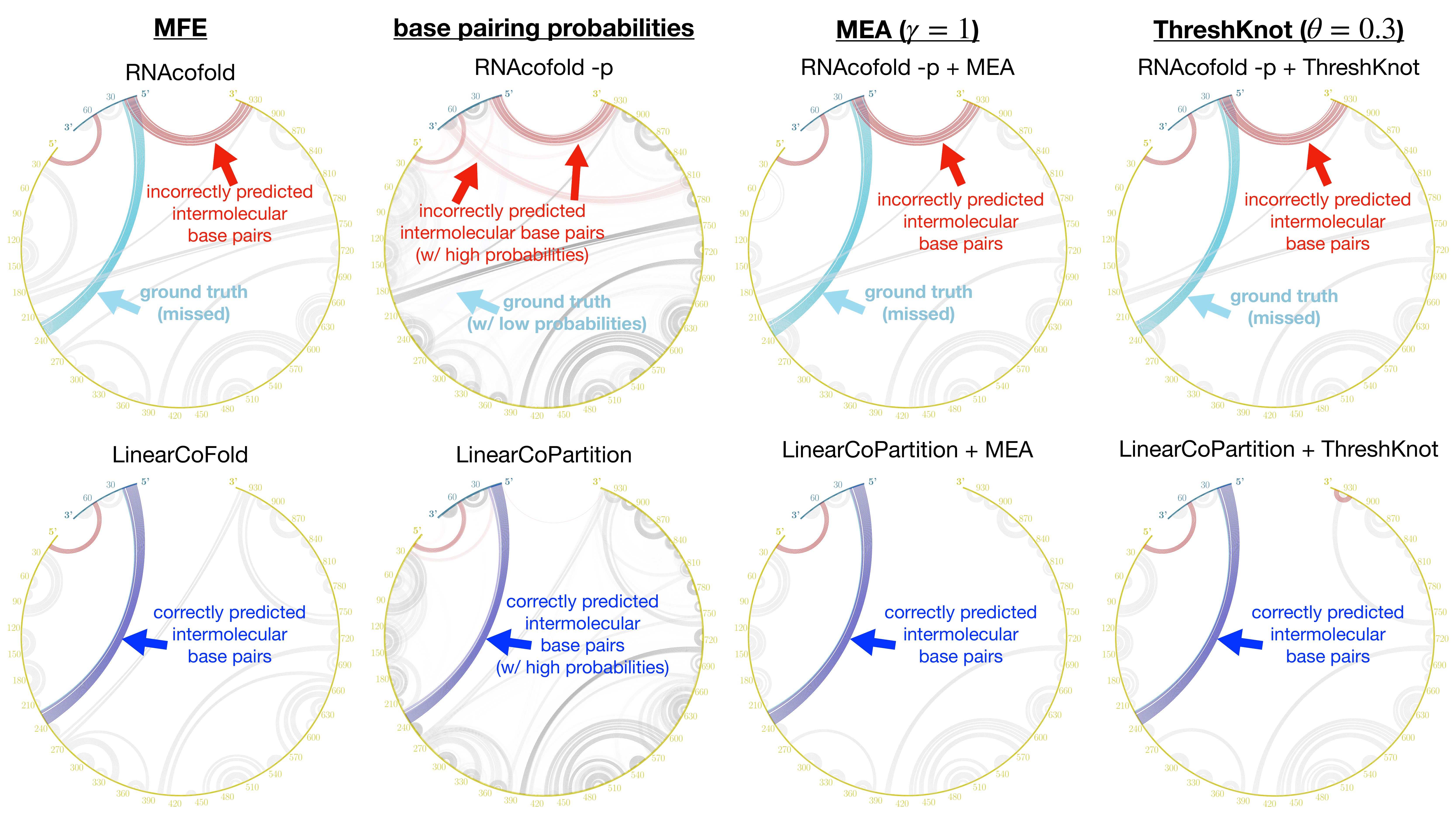} \\ [-9.2cm]
\begin{tabular}{cccc}
\hspace{-.0cm}\panel{D} & \hspace{3.75cm}\panel{E} & \hspace{3.65cm}\panel{F} & \hspace{3.55cm}\panel{G}\\[4.3cm]
\hspace{-.0cm}\panel{H} & \hspace{3.75cm}\panel{I} & \hspace{3.65cm}\panel{J} & \hspace{3.55cm}\panel{K}\\[-.0cm]
\end{tabular}
\vspace{4cm}
\caption{
Prediction accuracy comparison between \viennarnacofold and our algorithms on the Meyer dataset.
{\bf A}: PPV against sensitivity of the MFE structures (\rnacofold \crule[red]{.1cm}{.1cm} vs.~\lcf \crule[blue]{.1cm}{.1cm}), 
the MEA structures with varying $\gamma$ of 1, 1.5, 2 and 4 (\rnacofold $\filledcirc[red]$ vs.~\lcp $\filledcirc[blue]$), and the ThreshKnot structures with varying $\theta$ of 0.2, 0.25, 0.3 and 0.35 (\rnacofold $\textcolor{red} \triangle$ vs.~\lcp $\textcolor{blue} \triangle$).
{\bf B} and {\bf C}: per family and overall PPV and sensitivity comparisons between the six systems; we choose $\gamma=1$ for MEA and $\theta=0.3$ for ThreshKnot since they are the default values.
{\bf {D--K}}: circular plots of the MFE structure, the base pair probabilities, the MEA structure ($\gamma=1$) and the ThreshKnot structure ($\theta=0.3$) generated from \rnacofold ({\bf {D--G}}) and ours ({\bf {H--K}}) on a bacterial sRNA-mRNA sequence pair (MG1655 and NC\_000913.3), respectively; each arc represents a base pair (the darkness of the arc represents the pairing probability in {\bf E} and {\bf I}). 
The cyan arcs are the ground truth intermolecular base pairs;
the blue arcs are the correct predictions and
the red arcs are the incorrect predictions.
The intramolecular base pairs are colored in gray. 
	\label{fig:acc}
}
\end{figure*}

We compared the accuracy of \lcf and \lcp to \rnacofold  on the Meyer dataset.
Due to the absence of the annotation of intramolecular base pairs in the Meyer dataset, 
the accuracy evaluation is limited to intermolecular ones.
More specifically, 
we removed all intramolecular base pairs from the prediction, 
and calculated Positive Predictive Value (PPV, the fraction of predicted pairs in the annotated base pairs) 
and sensitivity (the fraction of annotated pairs predicted) 
to measure the accuracy only for intermolecular base pairs across the two families in the Meyer dataset,
and got the overall accuracy averaged on the two families.

Fig.~\ref{fig:acc}A shows the overall PPV and sensitivity on the Meyer dataset.
Compared to \rnacofold MFE mode, 
the overall PPV and sensitivity of \lcf increase 4.0\% and 11.6\%, respectively.
For the MEA structure prediction, 
we plotted a curve with varying $\gamma$ (a parameter balances PPV and sensitivity in the MEA algorithm) from 1 to 4;
compared to \rnacofold MEA, 
\lcf MEA shifts to the top-right corner,
which means that it has higher PPV and sensitivity.
For $\gamma=1$,
the overall PPV and sensitivity of \lcp MEA increase 2.9\% and 5.7\%, respectively.
In addition, for the ThreshKnot structures~\cite{Zhang+:2019}, 
we plotted a curve with varying $\theta$ (a parameter balances PPV and sensitivity in the ThreshKnot algorithm) from 0.2 to 0.35;
compared to \rnacofold ThreshKnot, 
\lcf ThreshKnot also shifts to the top-right corner.
For $\theta =0.3$,
the overall PPV and Sensitivity of \lcp ThreshKnot increase 2.4\% and 5.5\%, respectively.
Fig.~\ref{fig:acc}B and C show the PPV and sensitivity comparisons on each family,
which confirms that \lcf and \lcp are more accurate than \rnacofold on both bacterial sRNA-mRNA and fungal snoRNA-rRNA families.
 
On a bacterial sRNA-mRNA sequence pair (OmrA sRNA, 88~\nts; csgD mRNA, 951~\nts), 
we illustrated the MFE structures, 
the base-pairing probabilities, the MEA structures ($\gamma=1$) and the ThreshKnot structures ($\theta=0.3$) generated from \rnacofold MFE mode, partition-function (-p) mode, as well as \lcf and \lcp (Fig.~\ref{fig:acc}D--K). 
Each arc in the circular plots represents a base pair. 
The darkness of the arc represents its probability in the base-pairing matrix (Fig.~\ref{fig:acc}E and I). 
The intramolecular base pairs are in gray, 
while the intermolecular base pairs are marked using different colors to represent the correctly predicted pairs (blue),  
the ground-truth pairs but missing in the prediction (cyan),
and the incorrectly predicted pairs (red).  
We observed that all of our predictions correctly detect the intermolecular base pairs between 5'-end of the first strand and around 230~\nts of the second strand (blue arcs in Fig.~\ref{fig:acc}H--K),
while 
all of \rnacofold structures do not have these interactions (cyan arcs in Fig.~\ref{fig:acc}D--G), 
also incorrectly predict interactions between 5' end of the first strand and 3' end of the second strand (red arcs in Fig.~\ref{fig:acc}D--G). 

In \rnacofold,
the order of the two sequences does not matter, i.e., the predictions are the same when switching the two input sequences.
But in \lcf and \lcp, switching the order may result in different prediction, 
because the beam pruning heuristic may prune out different states when concatenating two strands in different orders.
We notice that \lcf and \lcp have higher accuracy on the Meyer dataset when using an oligo-first order 
(i.e., shorter sequence as the first input sequence and the longer one as the second).
This is because the Meyer dataset only annotates the intermolecular base pairs;
more intermolecular base pairs survive after beam pruning in the oligo-first order since there are less intramolecular base pairs competing with them. 
Therefore, we use the oligo-first order as default, and all results in Fig.~\ref{fig:acc} are in this order. 
We also present the accuracy of the reverse order on the Meyer dataset in Fig.~\ref{fig:reverse_order}.

\subsection{The prediction of host-virus RNA-RNA interaction}

\begin{figure*}[!hbt]
\hspace{.cm}\includegraphics[width=.99\textwidth]{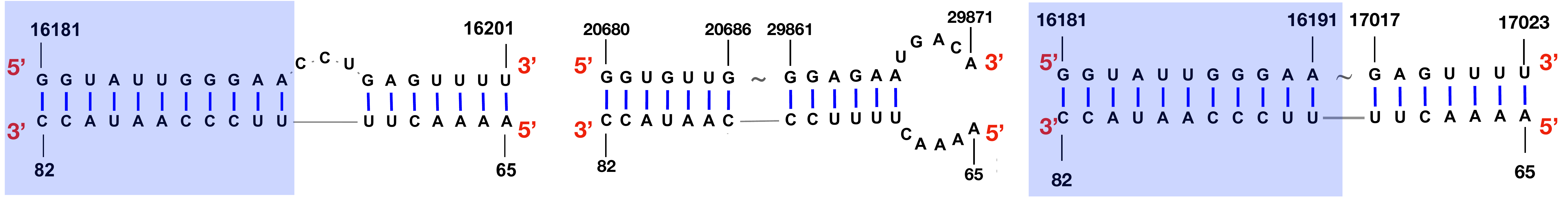} \\[-2.3cm]
\begin{tabular}{ccc}
\hspace{-4.6cm}\panel{A} & \hspace{-2.8cm}\panel{B} & \hspace{-4.1cm}\panel{C}\\[1.9cm]
\hspace{1.6cm}{\paneltext wetlab experiment} & \hspace{2.35cm}{\paneltext RNAcofold prediction} & \hspace{1.9cm}{\paneltext LinearCoFold prediction}
\end{tabular}
\caption{\lcf's prediction of the interaction between SARS-CoV-2 gRNA and human U4 snRNA better correlates with the wet lab experiments.
{\bf A}: the structure of SARS-CoV-2 gRNA and human snRNA U4 interacting region detected by the wet lab experiment.
{\bf B}: \rnacofold's prediction of the interacting structure.
{\bf C}: \lcf's prediction of the interacting structure.
The blue rectangles highlight the region that \lcf correlates with the wet lab experiment.
	\label{fig:covid}
}
\end{figure*}

Some viral genomes interact with the host RNAs.
A previous study~\cite{ziv2020short} found that the SARS-CoV-2 gRNA binds with human U4 small nuclear RNAs (snRNAs),
and illustrated their interacting structures, which are visualized in Fig.~\ref{fig:covid}A.
We can see that the [65, 82] region of human U4 snRNA forms helices with [16181, 16201] region of SARS-CoV-2 gRNA,
and a 3-nucleotide bulge loop locates in [16192, 16194] region.
Fig.~\ref{fig:covid}B shows that the predicted structure from \rnacofold does not match with the wet lab experiment results,
in which the [70, 82] region of human U4 snRNA pairs with the downstream region of SARS-CoV-2 gRNA.
By contrast, \lcf's prediction, shown in Fig.~\ref{fig:covid}C,
has intermolecular base pairs between [73, 82] region of human U4 snRNA and [16181, 16191] region of SARS-CoV-2 gRNA,
which overlaps with the experimental results 
and correctly predicts 11 out of 18 intermolecular base pairs.

\label{sec:discussion}

\section{Discussion}

\subsection{Summary}

We present \lcf and \lcp for the secondary structure prediction of two interacting RNA molecules.
Our two algorithms follow the strategy used \viennarnacofold,
which concatenates two RNA sequences and distinguishes ``normal loops'' from loops that contains nick point,
to simplify two-strand folding into the classical single-strand folding,
and predict both intramolecular and intermolecular interactions.
Based on this, \lcf and \lcp further apply beam pruning heuristics to reduce the cubic runtime in the classical RNA folding algorithms,
resulting in a linear-time prediction of minimum free energy structure (\lcf)
and a linear-time computation of partition function and base pairing probabilities (\lcp).
Unlike other {\it local} cofolding algorithms, 
\lcf and \lcp are {\it global} linear-time algorithms, 
which means that they do not have any limitations of base pairing distance,
allowing the prediction of global structures involving long distance interactions.
We confirm that:

\begin{enumerate}
\item \lcf and \lcp both run in linear time and space, and are orders of magnitude faster than \viennarnacofold.
On a sequence pair with combined length of 26,190~\nts,
\lcf is $86.8 \times$ faster than \rnacofold MFE mode,
and \lcp is $642.3 \times$ faster than \rnacofold partition function mode. 
See Fig.~\ref{fig:time_mem}. \\
\item Evaluated on the Meyer dataset with annotated intermolecular base pairs, 
\lcf and \lcp's predictions have higher PPV and sensitivity.
The overall PPV and Sensitivity of \lcf increase +4.0\% and +11.6\% over \rnacofold MFE, respectively;
\lcp MEA increases +2.9\% on PPV and +5.7\% on sensitivity over \rnacofold MEA,
and \lcp TheshKnot increases +2.4\% on PPV and +5.5\% on sensitivity over \rnacofold TheshKnot. 
See Fig.~\ref{fig:acc}A--C. 
A case study on a bacterial sRNA-mRNA sequence pair is provided to show the difference of predicted structures. 
See Fig.~\ref{fig:acc}D--K. \\
\item \lcf can predicts interaction between viral genomes and host RNAs.
For the SARS-CoV-2 gRNA interacting with human U4 snRNA confirmed by a previous wet lab study,
\lcf correctly predicts 11 out of 18 intermolecular base pairs,
while \rnacofold predicts 0 out of 18.
See Fig.~\ref{fig:covid}.\\
\end{enumerate}


\subsection{Extensions}

Our algorithm has several potential extensions.

\begin{enumerate}

\item 
Multiple RNAs can form into complex confirmation,
but current algorithms and tools are built on the classical $O(n^3)$ folding algorithms,
and are slow for long sequences~\cite{dirks2007thermodynamic}.
Our \lcf and \lcp are extendable from two-strand cofolding to multi-strand folding.

\item 
Following LinarSampling~\cite{zhang2020linearsampling}, a linear-time stochastic sampling algorithm for single strand, our \lcp is extendable to LinearCoSampling for the sampling of the cofolding structures.


\end{enumerate}




\acknow{\small 
}



\section*{References}
\bibliography{main}

\begin{thebibliography}{10}

\bibitem{tat2016cotranslational}
TT Tat, PA Maroney, S Chamnongpol, J Coller, TW Nilsen, {Cotranslational
  microRNA mediated messenger RNA destabilization}.
\newblock {\em\protect\JournalTitle{eLife}} \textbf{5}, e12880 (2016).

\bibitem{xu2016microrna}
K Xu, J Lin, R Zandi, JA Roth, L Ji, {MicroRNA-mediated target mRNA cleavage
  and 3'-uridylation in human cells}.
\newblock {\em\protect\JournalTitle{Scientific reports}} \textbf{6}, 1--14
  (2016).

\bibitem{rogers1980mechanism}
J Rogers, R Wall, {A mechanism for RNA splicing}.
\newblock {\em\protect\JournalTitle{Proceedings of the National Academy of
  Sciences}} \textbf{77}, 1877--1879 (1980).

\bibitem{mckeown1993role}
M McKeown, {The role of small nuclear RNAs in RNA splicing}.
\newblock {\em\protect\JournalTitle{Current opinion in cell biology}}
  \textbf{5}, 448--454 (1993).

\bibitem{kiss2002small}
T Kiss, Small nucleolar {RNA}s: an abundant group of noncoding {RNA}s with
  diverse cellular functions.
\newblock {\em\protect\JournalTitle{Cell}} \textbf{109}, 145--148 (2002).

\bibitem{elbashir2001duplexes}
SM Elbashir, et~al., Duplexes of 21-nucleotide {RNA}s mediate {RNA}
  interference in cultured mammalian cells.
\newblock {\em\protect\JournalTitle{Nature}} \textbf{411}, 494--498 (2001).

\bibitem{yuan2019approval}
H Yuan-Yu, Approval of the first-ever {RNA}i therapeutics and its technological
  development history.
\newblock {\em\protect\JournalTitle{Progress in Biochemistry and Biophysics}}
  \textbf{46}, 313--322 (2019).

\bibitem{hu2020therapeutic}
B Hu, et~al., Therapeutic si{RNA}: state of the art.
\newblock {\em\protect\JournalTitle{Signal transduction and targeted therapy}}
  \textbf{5}, 1--25 (2020).

\bibitem{stephenson1978inhibition}
ML Stephenson, PC Zamecnik, Inhibition of rous sarcoma viral {RNA} translation
  by a specific oligodeoxyribonucleotide.
\newblock {\em\protect\JournalTitle{Proceedings of the National Academy of
  Sciences}} \textbf{75}, 285--288 (1978).

\bibitem{dias2002antisense}
N Dias, C Stein, Antisense oligonucleotides: basic concepts and mechanisms.
\newblock {\em\protect\JournalTitle{Molecular cancer therapeutics}} \textbf{1},
  347--355 (2002).

\bibitem{rinaldi2018antisense}
C Rinaldi, MJ Wood, Antisense oligonucleotides: the next frontier for treatment
  of neurological disorders.
\newblock {\em\protect\JournalTitle{Nature Reviews Neurology}} \textbf{14},
  9--21 (2018).

\bibitem{wiedenheft2012rna}
B Wiedenheft, SH Sternberg, JA Doudna, {RNA}-guided genetic silencing systems
  in bacteria and archaea.
\newblock {\em\protect\JournalTitle{Nature}} \textbf{482}, 331--338 (2012).

\bibitem{zhang2018structural}
C Zhang, et~al., Structural basis for the {RNA}-guided ribonuclease activity of
  crispr-cas13d.
\newblock {\em\protect\JournalTitle{Cell}} \textbf{175}, 212--223 (2018).

\bibitem{bandaru2020structure}
S Bandaru, et~al., Structure-based design of g{RNA} for cas13.
\newblock {\em\protect\JournalTitle{Scientific reports}} \textbf{10}, 1--12
  (2020).

\bibitem{lorenz+:2011}
R Lorenz, et~al., Vienna{RNA} package 2.0.
\newblock {\em\protect\JournalTitle{Algorithms for Molecular Biology}}
  \textbf{6}, 1 (2011).

\bibitem{bernhart2011rna}
SH Bernhart, U M{\"u}ckstein, IL Hofacker, {RNA} accessibility in cubic time.
\newblock {\em\protect\JournalTitle{Algorithms for Molecular Biology}}
  \textbf{6}, 1--7 (2011).

\bibitem{mathews1999predicting}
DH Mathews, ME Burkard, SM Freier, JR Wyatt, DH Turner, Predicting
  oligonucleotide affinity to nucleic acid targets.
\newblock {\em\protect\JournalTitle{{RNA}}} \textbf{5}, 1458--1469 (1999).

\bibitem{rehmsmeier+:2004}
M Rehmsmeier, P Steffen, M Hochsmann, R Giegerich, Fast and effective
  prediction of micro{RNA}/target duplexes.
\newblock {\em\protect\JournalTitle{{RNA}}} \textbf{10}, 1507--1517 (2004).

\bibitem{tafer+hofacker:2008}
H Tafer, IL Hofacker, {RNA}plex: a fast tool for {RNA}–{RNA} interaction
  search.
\newblock {\em\protect\JournalTitle{Bioinformatics}} \textbf{24}, 2657--2663
  (2008).

\bibitem{muckstein2006thermodynamics}
U M{\"u}ckstein, et~al., Thermodynamics of {RNA}--{RNA} binding.
\newblock {\em\protect\JournalTitle{Bioinformatics}} \textbf{22}, 1177--1182
  (2006).

\bibitem{andronescu2005secondary}
M Andronescu, ZC Zhang, A Condon, Secondary structure prediction of interacting
  {RNA} molecules.
\newblock {\em\protect\JournalTitle{Journal of molecular biology}}
  \textbf{345}, 987--1001 (2005).

\bibitem{bernhart+:2006}
SH Bernhart, et~al., Partition function and base pairing probabilities of {RNA}
  heterodimers.
\newblock {\em\protect\JournalTitle{Algorithms for Molecular Biology}}
  \textbf{1} (2006).

\bibitem{Przybylska+:2009}
D Piekna-Przybylska, L DiChiacchio, DH Mathews, RA Bambara., A sequence similar
  to t{RNA}3lys gene is embedded in {HIV}-1 u3/r and promotes minus strand
  transfer.
\newblock {\em\protect\JournalTitle{Nat. Struct. Mol. Biol.}} \textbf{17},
  83–--89 (2009).

\bibitem{ding+lawrence:2003}
Y Ding, CE Lawrence, A statistical sampling algorithm for {RNA} secondary
  structure prediction.
\newblock {\em\protect\JournalTitle{Nucleic acids research}} \textbf{31},
  7280--7301 (2003).

\bibitem{lu2008efficient}
ZJ Lu, DH Mathews, Efficient {siRNA} selection using hybridization
  thermodynamics.
\newblock {\em\protect\JournalTitle{Nucleic acids research}} \textbf{36},
  640--647 (2008).

\bibitem{lai2016comprehensive}
D Lai, IM Meyer, A comprehensive comparison of general {RNA}--{RNA} interaction
  prediction methods.
\newblock {\em\protect\JournalTitle{Nucleic acids research}} \textbf{44},
  e61--e61 (2016).

\bibitem{umu+gardner:2017}
SU Umu, PP Gardner, A comprehensive benchmark of {RNA}-{RNA} interaction
  prediction tools for all domains of life.
\newblock {\em\protect\JournalTitle{Bioinformatics}} \textbf{33}, 988--996
  (2017).

\bibitem{DiChiacchio+:2016}
L DiChiacchio, MF Sloma, DH Mathews., Accessfold: predicting {RNA}-{RNA}
  interactions with consideration for competing self-structure.
\newblock {\em\protect\JournalTitle{Bioinformatics}} \textbf{32}, 1033–--1039
  (2016).

\bibitem{bernhart+:2006c}
SH Bernhart, et~al., Partition function and base pairing probabilities of {RNA}
  heterodimers.
\newblock {\em\protect\JournalTitle{Algorithms for Molecular Biology}}
  \textbf{1} (2006).

\bibitem{Dirks+:2007}
R Dirks, J Bois, J Schaeffer, E Winfree, N Pierce., Thermodynamic analysis of
  interacting nucleic acid strands.
\newblock {\em\protect\JournalTitle{SIAM Rev.}} \textbf{49}, 65--88 (2007).

\bibitem{do+:2006}
C Do, D Woods, S Batzoglou, {CONTRA}fold: {RNA} secondary structure prediction
  without physics-based models.
\newblock {\em\protect\JournalTitle{Bioinformatics}} \textbf{22}, e90--e98
  (2006).

\bibitem{Zhang+:2019}
L Zhang, H Zhang, DH Mathews, L Huang, Threshknot: Thresholded probknot for
  improved {RNA} secondary structure prediction.
\newblock {\em\protect\JournalTitle{bioRxiv}} (2019).

\bibitem{nussinov+jacobson:1980}
R Nussinov, AB Jacobson, Fast algorithm for predicting the secondary structure
  of single-stranded {RNA}.
\newblock {\em\protect\JournalTitle{Proceedings of the National Academy of
  Sciences}} \textbf{77}, 6309--6313 (1980).

\bibitem{zuker+stiegler:1981}
M Zuker, P Stiegler, Optimal computer folding of large {RNA} sequences using
  thermodynamics and auxiliary information.
\newblock {\em\protect\JournalTitle{Nucleic Acids Research}} \textbf{9},
  133--148 (1981).

\bibitem{xia+:1998}
T Xia, et~al., Thermodynamic parameters for an expanded nearest-neighbor model
  for formation of {RNA} duplexes with watson-crick base pairs.
\newblock {\em\protect\JournalTitle{Biochemistry}} \textbf{37}, 14719--14735
  (1998) PMID: 9778347.

\bibitem{Zuker+Sankoff:1984}
M Zuker, D Sankoff., {RNA} secondary structures and their prediction.
\newblock {\em\protect\JournalTitle{Bulletin of Mathematical Biology}}
  \textbf{46}, 591–--621 (1984).

\bibitem{Mathews+:1999}
DH Mathews, J Sabina, M Zuker, DH Turner, Expanded sequence dependence of
  thermodynamic parameters improves prediction of {RNA} secondary structure.
\newblock {\em\protect\JournalTitle{Journal of molecular biology}}
  \textbf{288}, 911--940 (1999).

\bibitem{Mathews+:2004}
DH Mathews, et~al., Incorporating chemical modification constraints into a
  dynamic programming algorithm for prediction of {RNA} secondary structure.
\newblock {\em\protect\JournalTitle{Proceedings of the National Academy of
  Sciences}} \textbf{101}, 7287--7292 (2004).

\bibitem{huang+:2019}
L Huang, et~al., {LinearFold: linear-time approximate {RNA} folding by 5'-to-3'
  dynamic programming and beam search}.
\newblock {\em\protect\JournalTitle{Bioinformatics}} \textbf{35}, i295--i304
  (2019).

\bibitem{zhang2020linearpartition}
H Zhang, L Zhang, DH Mathews, L Huang, {LinearPartition}: linear-time
  approximation of {RNA} folding partition function and base-pairing
  probabilities.
\newblock {\em\protect\JournalTitle{Bioinformatics}} \textbf{36}, i258--i267
  (2020).

\bibitem{agarwal2015predicting}
V Agarwal, GW Bell, JW Nam, DP Bartel, Predicting effective micro{RNA} target
  sites in mammalian m{RNA}s.
\newblock {\em\protect\JournalTitle{eLife}} \textbf{4}, e05005 (2015).

\bibitem{ziv2020short}
O Ziv, et~al., The short-and long-range {RNA}-{RNA} interactome of sars-cov-2.
\newblock {\em\protect\JournalTitle{Molecular cell}} \textbf{80}, 1067--1077
  (2020).

\bibitem{dirks2007thermodynamic}
RM Dirks, JS Bois, JM Schaeffer, E Winfree, NA Pierce, Thermodynamic analysis
  of interacting nucleic acid strands.
\newblock {\em\protect\JournalTitle{SIAM review}} \textbf{49}, 65--88 (2007).

\bibitem{zhang2020linearsampling}
H Zhang, L Zhang, S Li, DH Mathews, L Huang, {LazySampling and LinearSampling:
  Linear-time stochastic sampling of RNA secondary structure with applications
  to SARS-CoV-2}.
\newblock {\em\protect\JournalTitle{BioRxiv}} (2020).

\end{thebibliography}

\appendix

\newpage

\onecolumn
\newpage
  \begin{centering}
    \vspace*{1cm}
    
    \textbf{\Large Supporting Information}\\
    \vspace{0.5cm}
    \textbf{\Large \lcf and \lcp: Linear-Time Secondary Structure Prediction Algorithms of Interacting RNA molecules}\\
    \vspace{0.5cm}
    \textbf{\large
    He Zhang, Sizhen Li, Liang Zhang, David H.~Mathews and Liang Huang}
    \vspace{1.5cm}
    
  \end{centering}

\setcounter{figure}{0}
\renewcommand{\thefigure}{SI\,\arabic{figure}} 
\setcounter{table}{0}
\renewcommand{\thetable}{SI\,\arabic{table}}


  \algrenewcommand\algorithmicindent{0.5em}%
  \algnewcommand\algorithmicforeach{\textbf{for each}}
\algdef{S}[FOR]{ForEach}[1]{\algorithmicforeach\ #1\ \algorithmicdo}
\begin{figure}[!h]
\center
\begin{minipage}{0.8\textwidth}
\begin{algorithmic}[1]
  \newcommand{\INDSTATE}[1][1]{\State\hspace{#1\algorithmicindent}}
  \setstretch{1.05} 
\Function{LinearCoFold}{$\vecxa, \vecxb, b$} \Comment{$b$: beam size}
    \State $n \gets$ length of $\vecxa$ \Comment{$n$: first sequence length}
    \State $m \gets$ length of $\vecxb$ \Comment{$m$: second sequence length}
    \State $\vecx \gets$ $\vecxa \circ \vecxb$ \Comment{concatenate two sequences}
    \State $C \gets$ hash() \Comment{hash table: from span $[i,j]$ to $\Cf{i}{j}$}
    \State $\Cf{j}{j-1} \gets 0$ for all $j$ in $1...n+m$ \Comment{base cases} \label{line:base}
    \For{$j=1 ... n+m$} \label{line:dp_start}
        \ForEach {$i$ such that $[i,\,j-1]$ in $C$} \Comment{$O(b)$ iterations} 
            \State $\Cf{i}{j} \gets \Cf{i}{j-1} + \delta(\vecx,j) $ \Comment{\nskip} \label{line:skip_lcf}
            \If{$x_{i-1}x_j$ in \{AU, UA, CG, GC, GU, UG\}}  \label{line:pair}
                \ForEach{$k$ such that $[k,\,i-2]$ in $C$} \Comment{$O(b)$ iters} 
                    \If{$i-1 \leq n$ {\bf and} $j > n$} \Comment{intermolecular base pair} \label{line:start}
                        \State $\Cf{k}{j} \gets  \min (\Cf{k}{j}, \Cf{k}{i-2} + \Cf{i}{j-1} + \zeta(\vecx,i-1,j)$) \Comment{\pop} \label{line:pop_fake_lcf}
                    \Else \Comment{intramolecular base pair}
                        \State $\Cf{k}{j} \gets  \min (\Cf{k}{j}, \Cf{k}{i-2} + \Cf{i}{j-1} + \xi(\vecx,i-1,j)$) \Comment{\pop} \label{line:pop_normal_lcf}
                    \EndIf \label{line:end}
                \EndFor
            \EndIf
        \EndFor \label{line:dp_end}
        \State $\textsc {LinearCoFoldBeamprune}(C, j, b)$ \Comment{choose top $b$ out of $C(\cdot,j)$} \label{line:beamprune}
    \EndFor 
    \vspace{-0.1cm}
    \State \Return $C$ 
\EndFunction
\end{algorithmic}

\begin{algorithmic}[1]
  \newcommand{\INDSTATE}[1][1]{\State\hspace{#1\algorithmicindent}}
  \setstretch{1.2} 
\Function{LinearCoFoldBeamprune}{$C, j, b$}
    \State $\candidates \gets$ hash() \Comment{hash table: from candidate $i$ to score}
    \ForEach{$i$ such that $[i,j]$ in $C$}
        \State $\candidates[i] \gets \Cf{1}{i-1} + \Cf{i}{j}$ \Comment{use $\Cf{1}{i-1} $ as prefix score}
    \EndFor
    \State $\candidates \gets \textsc {SelectTopB}(candidates, b)$ \Comment{select top-$b$ states by score}
    \ForEach{$i$ such that $[i,j]$ in $C$}
        \If{key $i$ not in $candidates$}
            \State {\bf delete} $[i,j]$ from $C$ \Comment{prune low-scoring states}
        \EndIf
    \EndFor
\EndFunction
\end{algorithmic}
\end{minipage}
\caption{
Pseudocode of a simplified versions of the \lcf. 
Here we model hash tables following Python dictionaries, where $(i, j) \in C$ checks whether the key $(i, j)$ is in the hash $C$; 
this is needed to ensure linear runtime. 
Real \lcf system is much more involved, but the pseudocode illustrates the left-to-right partition function calculation idea using a Nussinov-like fashion.
\label{fig:algorithm_lcf}}
\vspace{-.3cm}
\end{figure}

 
\begin{figure}[!t]
\center
\begin{minipage}{0.8\textwidth}
\begin{algorithmic}[1]
  \newcommand{\INDSTATE}[1][1]{\State\hspace{#1\algorithmicindent}}
  \setstretch{1.05} 
\Function{LinearCoPartitionInside}{$\vecxa, \vecxb, b$}  \label{line:inside} \Comment{$b$: beam size}
    \State $n \gets$ length of $\vecxa$ \Comment{$n$: first sequence length}
    \State $m \gets$ length of $\vecxb$ \Comment{$m$: second sequence length}
    \State $\vecx \gets$ $\vecxa \circ \vecxb$ \Comment{concatenate two sequences}
    \State $Q \gets$ hash() \Comment{hash table: from span $[i,j]$ to $\Qf{i}{j}$}
    \State $\Qf{j}{j-1} \gets 1$ for all $j$ in $1...n+m$ \Comment{base cases} \label{line:base}
    \For{$j=1 ... n+m$}
    \ForEach {$i$ such that $[i,\,j-1]$ in $Q$} \Comment{$O(b)$ iterations}
            \State $\Qf{i}{j} \pluseq \Qf{i}{j-1} \cdot e^{-\frac{\delta(\vecx,j)}{RT}} $ \Comment{\nskip} \label{line:skip_lp}
            \If{$x_{i-1}x_j$ in \{AU, UA, CG, GC, GU, UG\}}  \label{line:pair}
                \ForEach{$k$ such that $[k,\,i-2]$ in $Q$} \Comment{$O(b)$ iters} 
                    \If{$i-1 \leq n$ {\bf and} $j > n$} \Comment{intermolecular base pair}
                        \State $\Qf{k}{j} \pluseq {\Qf{k}{i-2} \cdot \Qf{i}{j-1} \cdot e^{-\frac{\zeta(\vecx,i-1,j)}{RT}}} $ \Comment{\pop} \label{line:pop_fake}
                    \Else  \Comment{intramolecular base pair} 
                        \State $\Qf{k}{j} \pluseq {\Qf{k}{i-2} \cdot \Qf{i}{j-1} \cdot e^{-\frac{\xi(\vecx,i-1,j)}{RT}}} $ \Comment{\pop} \label{line:pop_normal}
                    \EndIf
                \EndFor
            \EndIf
        \EndFor
        \State $\textsc {LinearCoPartitionBeamprune}(Q, j, b)$ \Comment{choose top $b$ out of $Q(\cdot,j)$} \label{line:beamprune}
        \EndFor
        \vspace{-0.1cm}
    \State \Return $Q$ \Comment{partition function $Q(\vecx)=\Qf{1}{n}$}
\EndFunction
\end{algorithmic}

\begin{algorithmic}[1]
  \newcommand{\INDSTATE}[1][1]{\State\hspace{#1\algorithmicindent}}
  \setstretch{1.2} 
\Function{LinearCoPartitionOutside}{$\vecx, Q$}  \Comment{outside calculation}  \label{line:outside}
    \State $\Qhat \gets$ hash() \Comment{hash table: from span $[i,j]$ to $\Qhatf{i}{j}$: outside partition function}
    \State $p \gets$ hash() \Comment{hash table: from span $[i,j]$ to $p_{i,j}$: base-pairing probability}
    \State $\Qhatf{1}{n+m} \gets 1$ \Comment{base case}
    \State $Q_{\rm bp} \gets 1$ \Comment{temporary variable} \label{line:base}
    \For{$j=n+m$ {\bf down to } $1$}
        \ForEach {$i$ such that $[i,\,j-1]$ in $Q$}\smallskip
            \State $\Qhatf{i}{j-1} \pluseq \Qhatf{i}{j} \cdot e^{-\frac{\delta(\vecx,j)}{RT}} $ \Comment{\nskip} \label{line:skip}
            \If{$x_{i-1}x_j$ in \{AU, UA, CG, GC, GU, UG\}}  \label{line:pair}
                \ForEach{$k$ such that $[k,\,i-2]$ in $Q$}\smallskip
                    \If{$i-1 \leq n$ {\bf and} $j > n$} \Comment{intermolecular base pair}
                        \State $Q_{\rm bp} \gets e^{-\frac{\zeta(\vecx,i-1,j)}{RT}}$
                    \Else \Comment{intramolecular base pair}
                        \State $Q_{\rm bp} \gets e^{-\frac{\xi(\vecx,i-1,j)}{RT}}$
                    \EndIf
                    \State $\Qhatf{k}{i-2} \pluseq {\Qhatf{k}{j} \cdot \Qf{i}{j-1} \cdot Q_{bp}}$ \Comment{\pop: left} 
                        \State $\Qhatf{i}{j-1} \pluseq {\Qhatf{k}{j} \cdot \Qf{k}{i-2} \cdot Q_{bp}}$ \Comment{\pop: right} 
                        \State $p_{i-1,\,j} \pluseq \displaystyle\frac{\Qhatf{k}{j} \cdot \Qf{k}{i-2}  \cdot  Q_{bp} \cdot \Qf{i}{j-1}}{\Qf{1}{n+m}}$ \label{line:bpp} \Comment{accumulate base pairing probs}
                \EndFor
            \EndIf
        \EndFor
    \EndFor
    \State \Return $p$ \Comment{return the (sparse) base-pairing probability matrix}
\EndFunction
\end{algorithmic}

\begin{algorithmic}[1]
  \newcommand{\INDSTATE}[1][1]{\State\hspace{#1\algorithmicindent}}
  \setstretch{1.2} 
\Function{LinearCoPartitionBeamprune}{$Q, j, b$}
    \State $\candidates \gets$ hash() \Comment{hash table: from candidates $i$ to score}
    \ForEach{$i$ such that $[i,j]$ in $Q$}
        \State $\candidates[i] \gets \Qf{1}{i-1} \cdot \Qf{i}{j}$ \Comment{use $\Qf{1}{i-1} $ as prefix score}
    \EndFor
    \State $\candidates \gets \textsc {SelectTopB}(candidates, b)$ \Comment{select top-$b$ states by score} \label{line:quickselect}
    \ForEach{$i$ such that $[i,j]$ in $Q$}
        \If{key $i$ not in $candidates$}
            \State {\bf delete} $[i,j]$ from $Q$ \Comment{prune low-scoring states}
        \EndIf
    \EndFor
\EndFunction
\end{algorithmic}
\end{minipage}
\caption{
Pseudocode of a simplified versions of the \lcp,
including partition function calculation (inside phase) and base pairing probability calculation (outsite phase). 
\label{fig:algorithm_lcp}}
\vspace{-.5cm}
\end{figure}



\begin{figure*}
\begin{tabular}{cc}
\hspace{-8.4cm}\panel{A} & \hspace{-8.4cm}\panel{B} \\[-.5cm]
\includegraphics[width=0.48\textwidth]{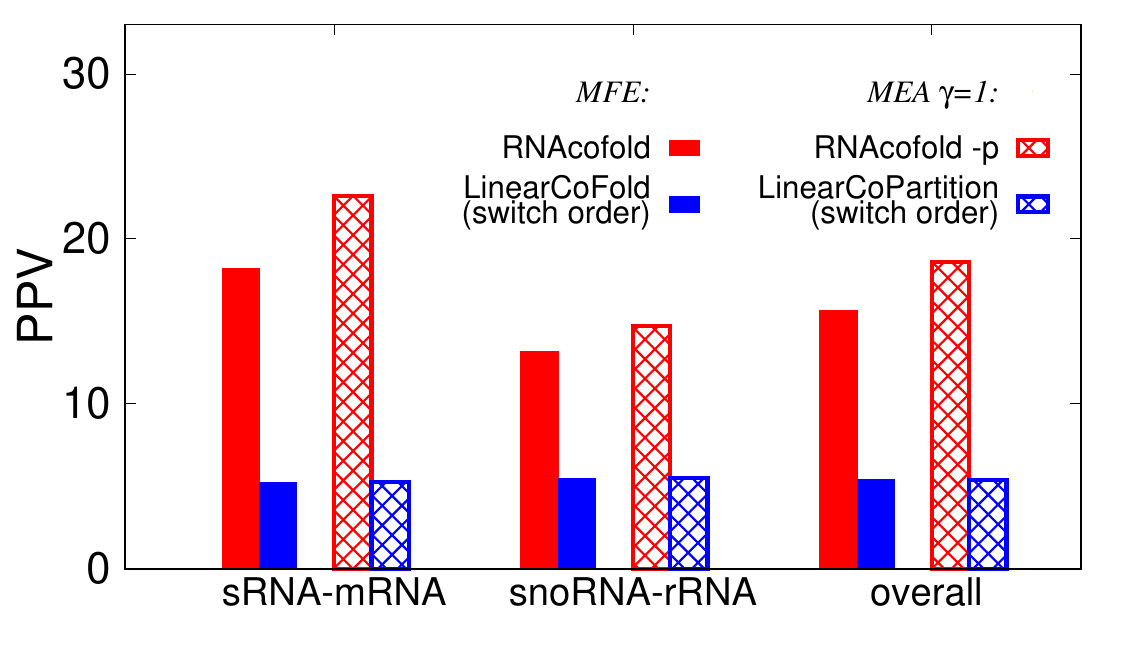} &
\includegraphics[width=0.48\textwidth]{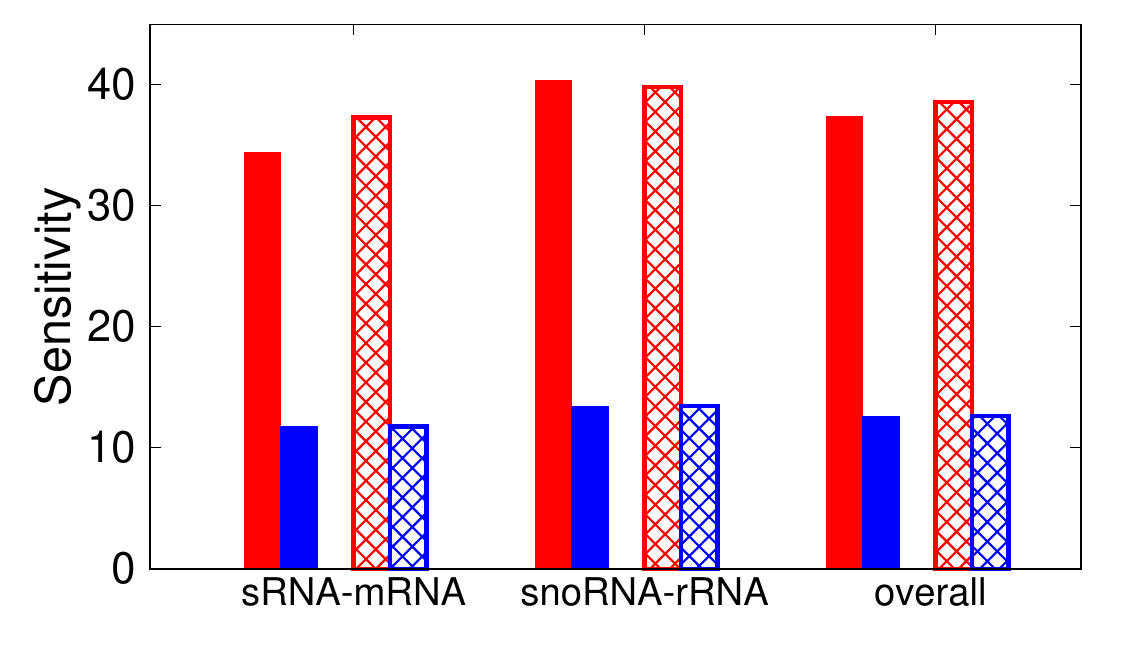} \\
\end{tabular}
\caption{The accuracies of \lcf and \lcp drop when switching the order of the two input sequences, i.e., longer sequence as the first input sequence and shorter sequence as the second one.
	\label{fig:reverse_order}
}
\end{figure*}

\end{document}